\documentclass[11pt,reqno]{amsproc}
\usepackage[T1]{fontenc}
\usepackage{amsmath,amscd,amssymb}
\usepackage{cite}
\usepackage{color}
\usepackage{graphicx}
\usepackage{psfrag,epsfig}
\usepackage{multirow}
\usepackage{rotating}
\usepackage[letterpaper,margin=0.5in]{geometry}
\usepackage{subfigure}
\usepackage{enumerate}
\usepackage{comment}
\usepackage{lineno}
\usepackage{algorithmic}
\usepackage{algorithm}
\usepackage[scriptsize]{caption}
\usepackage{bbold}
\usepackage[debug=false, colorlinks=true, pdfstartview=FitV, linkcolor=blue, citecolor=blue, urlcolor=blue]{hyperref}

\newcommand*{\Comb}[2]{{}^{#1}C_{#2}}%
\numberwithin{equation}{section}
\usepackage{chemarr}
\usepackage[version=3]{mhchem}
\newcommand*\patchAmsMathEnvironmentForLineno[1]{%
  \expandafter\let\csname old#1\expandafter\endcsname\csname #1\endcsname
  \expandafter\let\csname oldend#1\expandafter\endcsname\csname end#1\endcsname
  \renewenvironment{#1}%
     {\linenomath\csname old#1\endcsname}%
     {\csname oldend#1\endcsname\endlinenomath}}%
\newcommand*\patchBothAmsMathEnvironmentsForLineno[1]{%
  \patchAmsMathEnvironmentForLineno{#1}%
  \patchAmsMathEnvironmentForLineno{#1*}}%
\AtBeginDocument{%
\patchBothAmsMathEnvironmentsForLineno{equation}%
\patchBothAmsMathEnvironmentsForLineno{align}%
\patchBothAmsMathEnvironmentsForLineno{flalign}%
\patchBothAmsMathEnvironmentsForLineno{alignat}%
\patchBothAmsMathEnvironmentsForLineno{gather}%
\patchBothAmsMathEnvironmentsForLineno{multline}%
}
\usepackage{siunitx}
\DeclareSIUnit{\mEq}{mEq}
\title[Deep Learning for Permeability Inversion]{Deep Learning to Estimate Permeability using Geophysical Data}
\author[M.~K.~Mudunuru et.al.,]{M.~K.~Mudunuru$^{1,*}$, E.~L.~D.~Cromwell$^{2,*}$,  H.~Wang$^{3}$, and X.~Chen$^{1}$ \\
{\small $^{1}$Watershed \& Ecosystem Science, Pacific Northwest National Laboratory, Richland, WA 99352, USA} \\
{\small $^{2}$Artificial Intelligence \& Data, Pacific Northwest National Laboratory, Richland, WA 99352, USA} \\
{\small $^{3}$Department of Mining and Minerals Engineering, Virginia Tech, Blacksburg, VA, 24061, USA}}
\thanks{$^*$These authors contributed equally to this work. 
Corresponding author's email address:~\texttt{maruti@pnnl.gov}}
\date{\today}
\begin{document}
\maketitle
%
%
\section*{ABSTRACT}
Time-lapse electrical resistivity tomography (ERT) is a popular geophysical method to estimate three-dimensional (3D) permeability fields from electrical potential difference measurements.
Traditional inversion and data assimilation methods are used to ingest this ERT data into hydrogeophysical models to estimate permeability. 
Due to ill-posedness and the curse of dimensionality, existing inversion strategies provide poor estimates and low resolution of the 3D permeability field.
Recent advances in deep learning provide us with powerful algorithms to overcome this challenge.
This paper presents a deep learning (DL) framework to estimate the 3D subsurface permeability from time-lapse ERT data.
To test the feasibility of the proposed framework, we train DL-enabled inverse models on simulation data.
{\color{black}Each measurement in both synthetic and field data is standardized by removing the mean and scaling the time-series to unit variance.
This pre-processing step is necessary to bring simulation data closer to field observations.
Subsurface process models based on hydrogeophysics are used to generate this synthetic data.}
Training performed on limited simulation data resulted in the DL model over-fitting.
An advanced data augmentation based on mixup is implemented to generate additional training samples to overcome this issue.
This mixup technique creates weakly labeled (low-fidelity) samples from strongly labeled (high-fidelity) data.
The weakly labeled training data is then used to develop DL-enabled inverse models and reduce over-fitting.
{\color{black}As both time-lapse ERT (1133048 features/realization) and 3D permeability (585453 features/realization) data samples are from a high-dimensional space, principal component analysis (PCA) is employed to reduce dimensionality.
Encoded ERT and encoded permeability are generated using the trained PCA estimators.}
A deep neural network is then trained to map the encoded ERT to encoded permeability.
This mixup training and unsupervised learning allowed us to build a fast and reasonably accurate DL-based inverse model under limited simulation data.
Results show that proposed weak supervised learning can capture salient spatial features in the 3D permeability field.
Quantitatively, the average mean squared error (in terms of the natural log) on the strongly labeled training, validation, and test datasets is less than 0.5.
The $R^2$-score (global metric) is greater than 0.75, and the percent error in each cell (local metric) is less than 10\%. 
Finally, an added benefit in terms of computational cost is that the proposed DL-based inverse model is at least $\mathcal{O}(10^4)$ times faster than running a forward model once it is trained.
{\color{black}Data generation, DL model training, and hyperparameter tuning to identify optimal neural network architectures utilized high-performance computing resources while the DL inference is performed on a standard laptop.
Approximately, $\mathcal{O}(10^5)$ processor hours are used for generating data and DL tuning and training.
We acknowledge that the data generation and DL model development are expensive.
But once a DL model is trained, it can be re-used for inversion rapidly for the given system, with set physics and domain.}
Note that traditional inversion may require multiple forward model simulations (e.g., in the order of 10 to 1000), which are very expensive.
This computational savings $\left(\approx \mathcal{O}(10^5) - \mathcal{O}(10^7) \right)$ makes the proposed DL-based inverse model attractive for subsurface imaging and real-time ERT monitoring applications due to fast and yet reasonably accurate estimations of permeability field.
\\
\\
\textbf{KEYWORDS:}~Hydrogeophysics,
electric resistivity tomography,
permeability,
multi-physics,
dimensionality reduction,
deep neural networks,
scalable deep learning.

\section*{\textbf{HIGHLIGHTS AND NOVELTY}}
\begin{itemize}
    \item A deep learning (DL) framework is developed to estimate the 3D permeability using geophysical data.
    \item Mixup-based data augmentation is used to reduce over-fitting and improve DL-enabled inversion under limited training data. 
    \item Our tuned DL models capture salient structural features in the 3D permeability field with reasonable accuracy (e.g., $< 10\%$ error in each grid cell).
    \item An added benefit is that the proposed DL methodology is at least $\mathcal{O}(10^4)$ times faster than running a single simulation.
    {\color{black}\item However, the simulation data generation and DL model training is computationally expensive, which is in $\mathcal{O}(10^5)$ processor hours.}
\end{itemize}

\section{\textbf{INTRODUCTION}}
\label{Sec:S1_Intro}
Permeability plays a crucial role in the predictive understanding of subsurface systems \cite{al2004hydro,atekwana2009biogeophysics,chen2020integrating,mudunuru2021subsurface}.
Characterizing spatially distributed permeability with reasonably good accuracy is a significant challenge in the Earth and energy sciences \cite{berkowitz2002characterizing,neuman2005trends}.
Good accuracy is needed because fluid flow and reactive-transport in applications such as hydrology \cite{chen2020integrating}, hydro-biogeochemistry \cite{allen2007microbial,atekwana2009biogeophysics,atekwana2010geophysical}, unconventional oil and gas \cite{middleton2015shale,hyman2016understanding,mudunuru2020physics}, $\mathrm{CO}_2$ sequestration \cite{stauffer2009system,middleton2012cross,ahmmed2021machine}, nuclear waste disposal and storage \cite{rutqvist2009comparative,apted2017geological}, and geothermal systems are controlled by permeability \cite{brown2012mining,mudunuru2017regression,siler2021machine}.
Due to the subsurface's heterogeneous nature, much uncertainty is involved in estimating the three-dimensional (3D) permeability field. 
Hence, capturing critical structural features in permeability is essential for developing realistic subsurface process models for natural and engineered systems \cite{chen2020integrating,SMART_CS_Initiative}.

There are various subsurface imaging techniques to collect data for estimating 3D permeability field \cite{guerin2005borehole,turk2011subsurface,saleh2011introduction}.
Direct techniques (e.g., invasive imaging, well/core logging) generally sample measurements at different spatial locations and are point-wise \cite{turk2011subsurface,preko2009comparison,binley2015emergence}.
They provide high-resolution data only for small spatial coverage of the permeability field but are challenging, time-consuming, and very expensive to acquire.
{\color{black}Non-invasive techniques (e.g., geophysical tomographic methods such as seismic, electrical, magnetic, nuclear magnetic resonance) provide indirect data to estimate permeability \cite{binley2015emergence,singha2015advances,parsekian2015multiscale,misra2019deep,li2020neural}.}
This indirect measurement data is assimilated into process models (e.g., through parameter estimation algorithms) to estimate permeability.
A significant advantage of such non-invasive subsurface imaging techniques is that data collection at field scales is affordable.
However, the collected data can be noisy and may have a low spatial resolution.
Prominent non-invasive methods include electrical resistivity tomography, ground-penetrating radar, induced polarization, seismic, gravity, electromagnetics, and magnetotellurics \cite{slater2007near,robinson2008advancing,atekwana2009biogeophysics}.
Time-lapse electrical resistivity tomography (ERT) is a popular non-invasive geophysical method to estimate the permeability field from electrical potential difference measurements \cite{rubin2006hydrogeophysics}.
In this ERT technique, surface electrodes form an imaging array.
Surveys (e.g., given an array layout and measurement sequence) are conducted to collect time-lapse ERT data.
Each survey consists of thousands to tens of thousands of measurements (e.g., our study has 40,466 measurements, see Sec.~\ref{SubSec:S2_Prob_Description}).
The primary time-lapse ERT measurement is transfer resistance at different times.
Transfer resistance is the ratio of the observed electrical potential difference between a pair of electrodes to the injected current.
One such measurement requires a four-electrode configuration, which consists of a current source and sink electrode pair and a positive and negative electrical potential electrode pair.
The collected measurements from a time-lapse ERT survey are used to estimate subsurface permeability using inversion algorithms \cite{seo2012nonlinear,mueller2012linear,sen2013global}.

State-of-the-art inversion (e.g., Tikhonov, total variation, and Besov norm regularizations) \cite{tarantola2005inverse,aster2018parameter} and data assimilation (e.g., ensemble smoother, ensemble Kalman filers and its variants) \cite{chen2013application,asch2016data,jiang2021dart} are generally used to ingest the ERT data for estimating permeability.
Software packages such as \texttt{PEST} \cite{doherty2010approaches}, \texttt{DART} \cite{anderson2009data}, and \texttt{DAKOTA} \cite{adams2009dakota} can also be used to estimate permeability.
These tools and associated algorithms interact with a numerical model (e.g., discretizations of subsurface process models) to estimate and adjust process parameters based on the assimilated data \cite{jiang2021dart}.
However, there are certain limitations with the methods and tools mentioned earlier.
For instance, these data assimilation methods and software packages may require multiple forward model runs \cite{stuart2010inverse,mudunuru2017sequential,maclaren2019can}.
But high-fidelity process model simulations are computationally expensive, which require high-performance computing (HPC) resources
(see Sec.~\ref{SubSec:S4_Comp_Cost}).
These resources are not always available or accessible to perform detailed uncertainty quantification studies (e.g., calibration-constrained uncertainty analysis, prior and posterior uncertainty analysis, local and global sensitivity analysis \cite{caers2011modeling,scheidt2018quantifying,hu2008multiple}).
Moreover, due to ill-posedness and due to subsurface processes' nonlinearity, existing inversion algorithms provide poor estimates and low resolution of the process model parameters in high-dimensional space \cite{oware2013physically}.
Inverse modeling strategies that are fast and provide reasonably accurate estimations of high-dimensional parameters (e.g., in the $\mathcal{O}(10^2) - \mathcal{O}(10^6)$) are needed to overcome the above limitations.
Recent advances in machine learning (ML) and deep learning (DL) show promise to develop such models \cite{sun2019can,tahmasebi2020machine,jagtap2021deep,xu2021machine,cromwell2021estimating} once properly trained.

ML and DL algorithms based on neural networks provide new approaches to discover and map nonlinear relationships between process model inputs and outputs \cite{sun2018discovering,misra2019machine,camps2021deep,jagtap2021deep,cromwell2021estimating}.
They also provide cost-effective and efficient methods to transform high-dimensional data into low-dimensional space (e.g., through principal component analysis\cite{ringner2008principal,bro2014principal,siler20213}, matrix/tensor factorization \cite{cichocki2009nonnegative,vesselinov2019unsupervised}).
{\color{black}This dimensionality reduction allows us to extract hidden features or latent variables from high-dimensional data, thereby better representing the underlying processes and system response \cite{gonzalez2022monitoring}.}
However, training DL models (e.g., using deep neural networks) requires large amounts of training data \cite{sun2019can}.
Generating high-quality data can be very expensive and time-consuming \cite{jagtap2021deep,cromwell2021estimating}.
{\color{black}Recent advances in data augmentation\cite{shorten2019survey} (e.g., mixup \cite{zhang2017mixup}, neural style transfer \cite{jing2019neural}), transfer learning \cite{tan2018survey}, and meta-learning (e.g., zero-shot, one-shot, few-shot, and any-shot) \cite{finn2017model,vanschoren2018meta} are used to overcome this challenge of training DL models under limited sample size (e.g., in the $\mathcal{O}(10^2)$).}
In this study, we use mixup to generate additional samples during the training process.
Mixup-based data augmentation \cite{zhang2017mixup,liang2018understanding,thulasidasan2019mixup,zhang2020does,deng2021deep} is a simple strategy, which effectively produces weakly labeled data from high-quality samples to improve the predictive capability of DL models.
These inexpensive, weakly generated data are employed to train our DL models.
Hence, weak supervised learning based on mixup allows us to reduce model overfitting (e.g., when training under limited samples) \cite{zhang2020does}.

\subsection{\textbf{Main contributions}}
\label{SubSec:Main_Contribution}
The main contribution of this study is to develop fast and accurate deep learning models to estimate spatially distributed subsurface permeability from time-lapse ERT.
The DL-based inverse models are built on numerical simulations that comply with underlying hydrogeophysics processes.
The simulation data for training is developed by coupling flow, tracer-transport, and ERT-based geophysics process models.
An advantage of the proposed methodology is that it provides good accuracy (e.g., prediction error in each grid cell is less than 10\%) in estimating permeability.
We emphasize that improved accuracy is achieved by data augmentation through mixup.
An added benefit of the proposed DL-based inverse models is that it is at least $\mathcal{O}(10^4)$ times faster than running a forward model simulation\footnote{We acknowledge that generating simulation data is expensive and is generally required in large amounts for training a reliable DL model.}. 
From a computational cost perspective, traditional inversion requires multiple forward model runs.
As a result, inversion may demand HPC resources and can be prohibitively expensive.
Due to the savings in computational cost, our DL-based model inversion is attractive for usage in comprehensive uncertainty quantification studies towards real-time subsurface imaging.

\subsection{\textbf{Outline of the paper}}
\label{SubSec:Outline}
The paper is organized as follows: Sec.~\ref{Sec:S1_Intro} discusses the state-of-the-art model inversion methods.
It provides limitations of these methods along with the advantages of using DL for model inversion. 
Then, we present the need for a DL-enabled inversion framework for estimating the 3D permeability field from electrical potential measurements.
Sec.~\ref{Sec:S2_HGP_Model} describes the hydrogeophysics process models used to generate ERT training samples for permeability inversion.
Problem description and data generation statistics are also described.
Sec.~\ref{Sec:S3_DL_Framework} introduces the proposed deep learning-based model inversion framework.
DL models are trained under limited high-quality training samples.
Mixup training samples are used to reduce model overfitting.
Deep learning model architecture, associated training process, and hyperparameter tuning are also discussed.
Results are presented in Sec.~\ref{Sec:S4_Results}.
The computational cost of generating data, running forward models, and DL-based inverse models is described.
We also discuss the limitations of the proposed method and possible approaches to improve it.
Finally, conclusions are drawn in Sec.~\ref{Sec:S5_Conclusions}.

\section{\textbf{HYDROGEOPHYSICS PROCESS MODELS AND TRAINING DATA GENERATION}}
\label{Sec:S2_HGP_Model}
In this section, the hydrogeophysics process models \cite{johnson2017pflotran,ahmmed2020pflotran} used to generate time-lapse ERT simulations are presented first.
We couple subsurface flow, tracer transport, and geoelectrical process models to develop training samples for DL.
\texttt{PFLOTRAN} \cite{hammond2012pflotran,Hammondetal2014,lichtner2015pflotran}, an open-source multi-physics code is used to simulate subsurface flow and tracer transport.
\texttt{E4D} \cite{johnsonetal2010,johnson2017pflotran}, a geophysics code is used to simulate time-lapse ERT.
These two codes are loosely coupled using Jupyter notebooks.
{\color{black}By loosely coupled, we mean \texttt{PFLOTRAN} simulations are performed first.
Then, the outputs generated from PFLOTRAN (e.g., concentration, saturation) are transferred to \texttt{E4D}'s mesh to run time-lapse ERT simulations.
This entire workflow of coupling \texttt{PFLOTRAN} and \texttt{E4D} is performed using Jupyter notebooks.}
Text S1 in the supplementary material describes {\color{black}the process models} to develop simulation data.
Specifics on the modeling domain that represent the field study of stage-driven groundwater/river water interaction along the Columbia River, Washington, USA, are outlined.
Finally, the workflow to generate the samples for training the DL-based inverse models is described.

\begin{figure}[!htbp]
  \centering
    \subfigure[\texttt{PFLOTRAN} and \texttt{E4D} modeling domain]
    {\includegraphics[width = 0.55\textwidth]{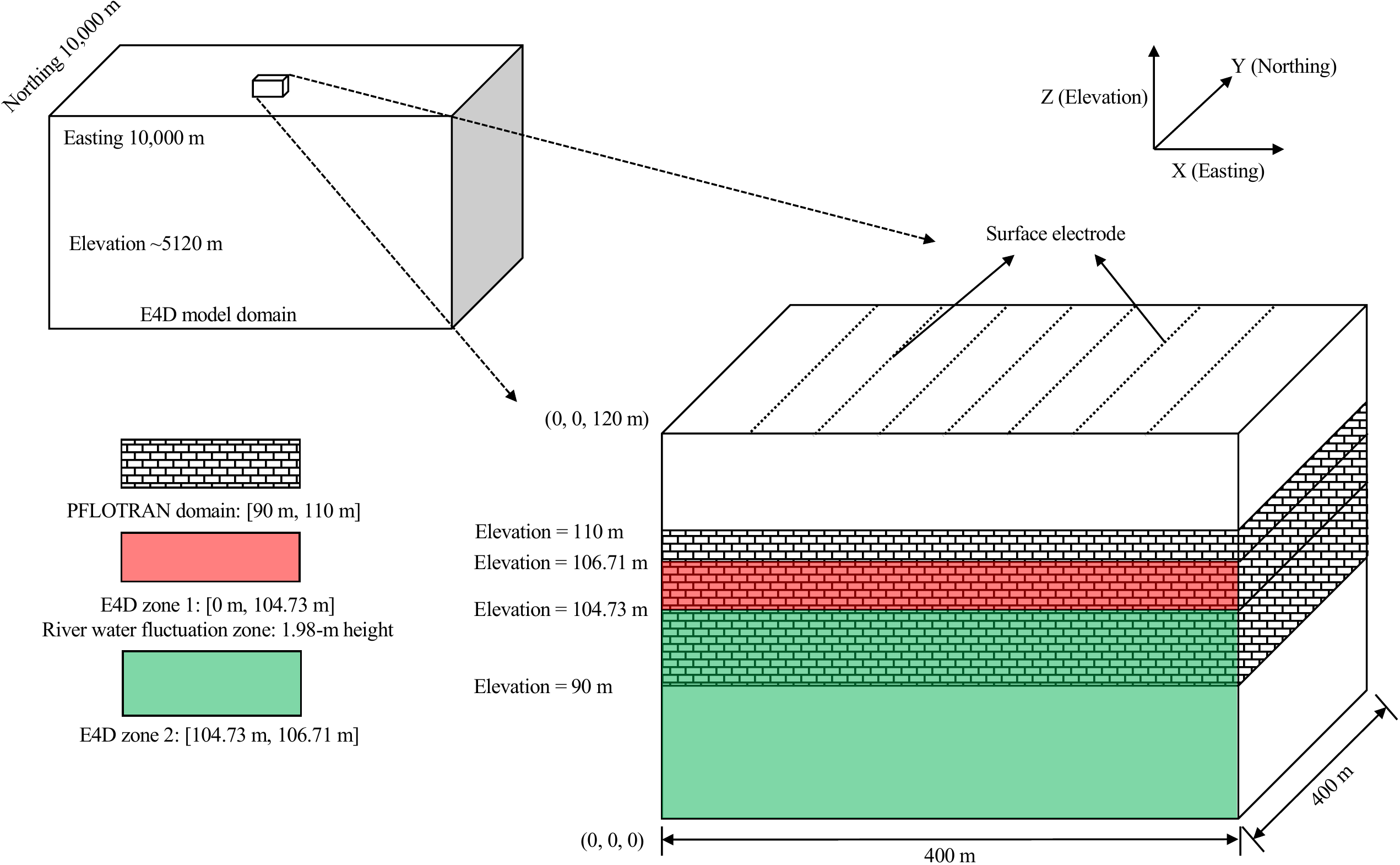}}
    \subfigure[\texttt{E4D} mesh and refined electrode area]
    {\includegraphics[width = 0.55\textwidth]{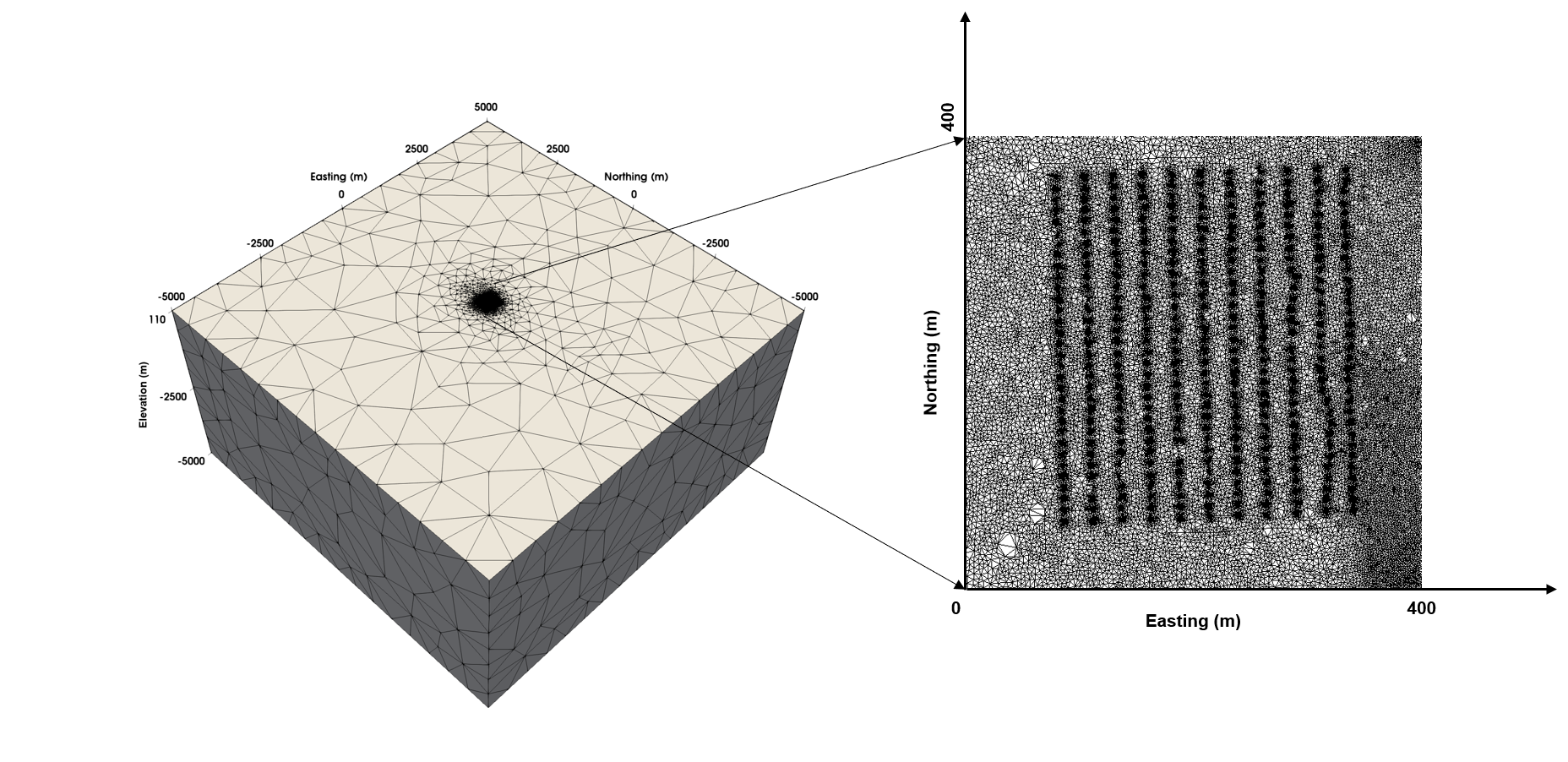}}
  \caption{\textbf{Problem description:}~A pictorial description of boundary value problem.
  The top figure shows the modeling domains for simulating \texttt{PFLOTRAN} and \texttt{E4D} process models.
  The bottom figure shows the unstructured tetrahedral mesh and refined electrode area in the \texttt{E4D} modeling domain.
  \label{Fig:Problem_Description}}
\end{figure}

\subsection{\textbf{Problem description}}
\label{SubSec:S2_Prob_Description}
Figure~\ref{Fig:Problem_Description} provides a pictorial description of the \texttt{PFLOTRAN} and \texttt{E4D} modeling domain.
This depiction is a representation of field study to monitor groundwater/surface-water interactions along the Columbia River at the Hanford 300 area in Washington, USA \cite{johnson2017pflotran}.
The study site is characterized by dynamic stage variations in the Columbia River, resulting in a variably saturated subsurface.
A 3D ERT monitoring array was installed to monitor the river water intrusion during April 2013 \cite[Section-3]{johnson2015four}.
Figure~\ref{Fig:Problem_Description}(b) shows the location of electrodes in a relative coordinate system.
It also provides an unstructured tetrahedral mesh employed by \texttt{E4D} to simulate potential difference measurements at the surface electrodes.
The \texttt{PFLOTRAN} and \texttt{E4D} modeling specifics are described in supplementary material text S2.

\subsection{\textbf{Training data generation}}
\label{SubSec:S2_Training_Data_Generation}
\begin{figure}[!htbp]
  \centering
  \subfigure[Training realization]
    {\includegraphics[width = 0.325\textwidth]
    {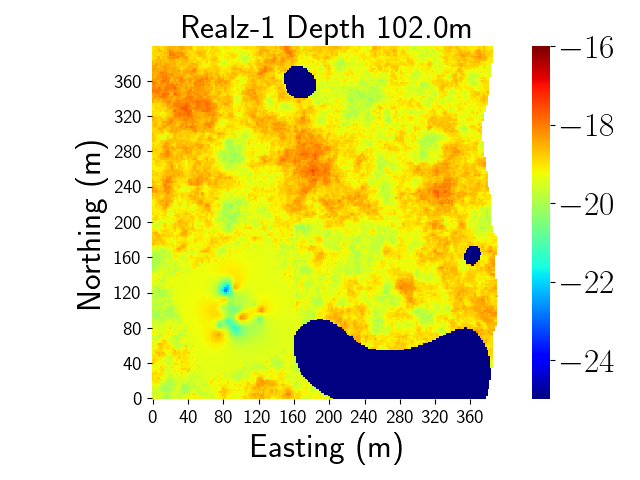}}
  \subfigure[Training realization]
    {\includegraphics[width = 0.325\textwidth]
    {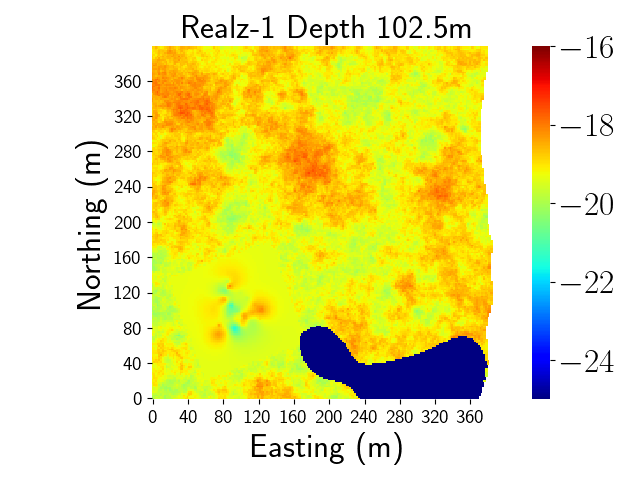}}
  \subfigure[Training realization]
    {\includegraphics[width = 0.325\textwidth]
    {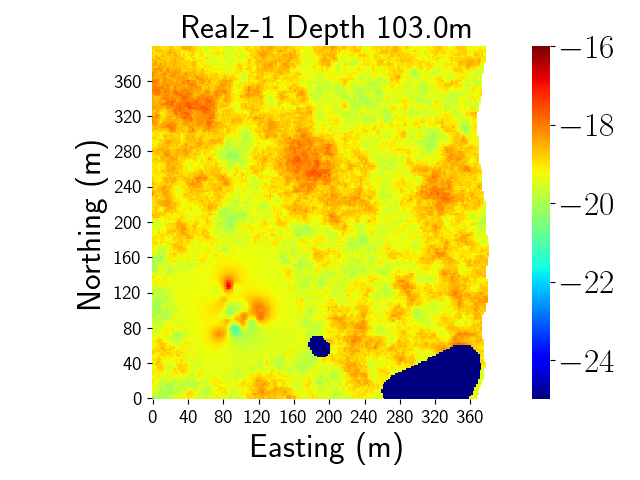}}
  \subfigure[Validation realization]
    {\includegraphics[width = 0.325\textwidth]
    {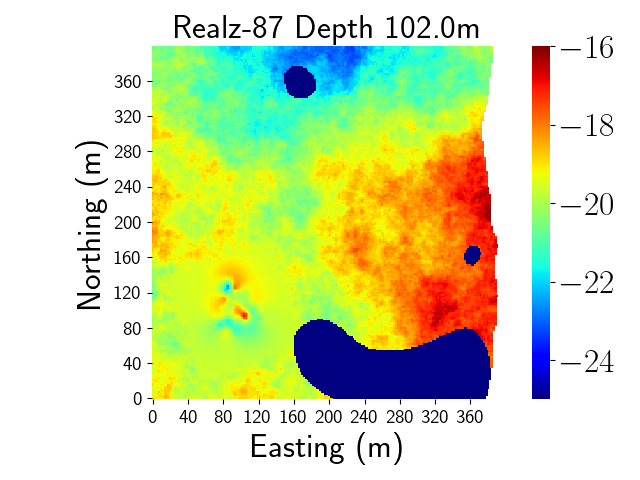}}
  \subfigure[Validation realization]
    {\includegraphics[width = 0.325\textwidth]
    {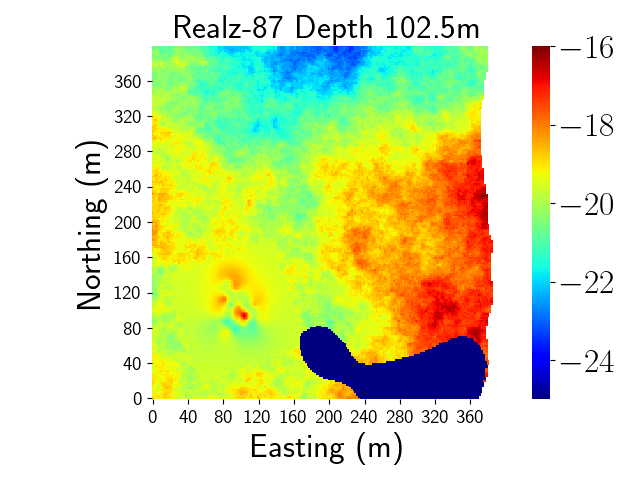}}
  \subfigure[Validation realization]
    {\includegraphics[width = 0.325\textwidth]
    {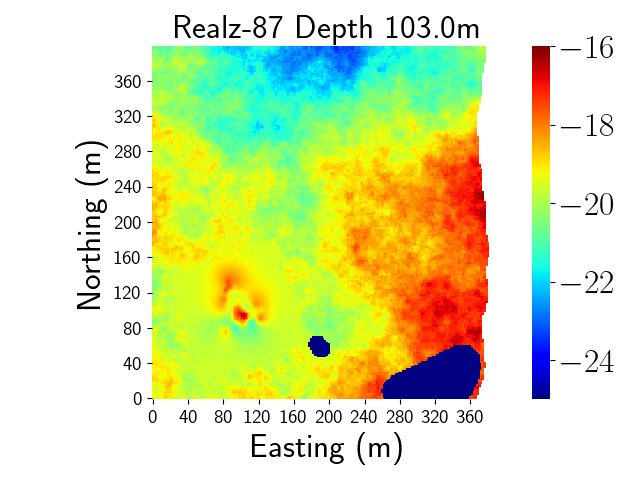}}
    \subfigure[Test realization]
    {\includegraphics[width = 0.325\textwidth]
    {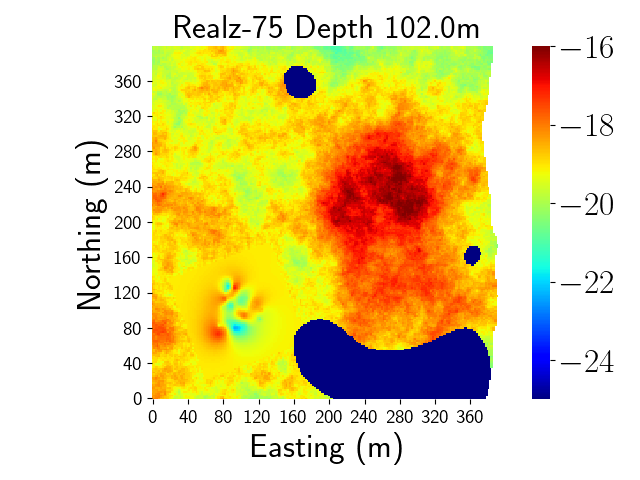}}
  \subfigure[Test realization]
    {\includegraphics[width = 0.325\textwidth]
    {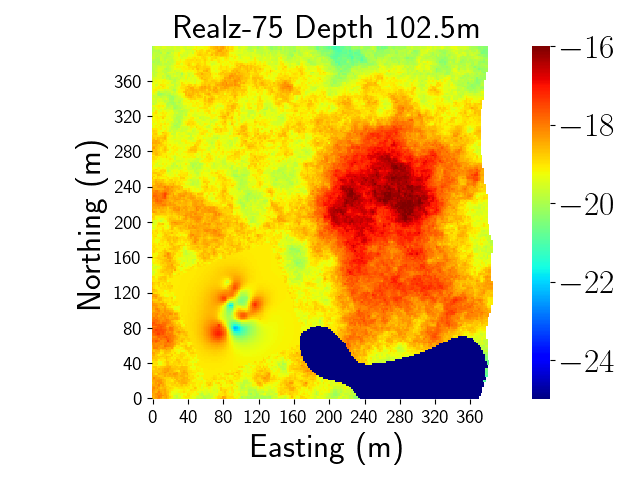}}
  \subfigure[Test realization]
    {\includegraphics[width = 0.325\textwidth]
    {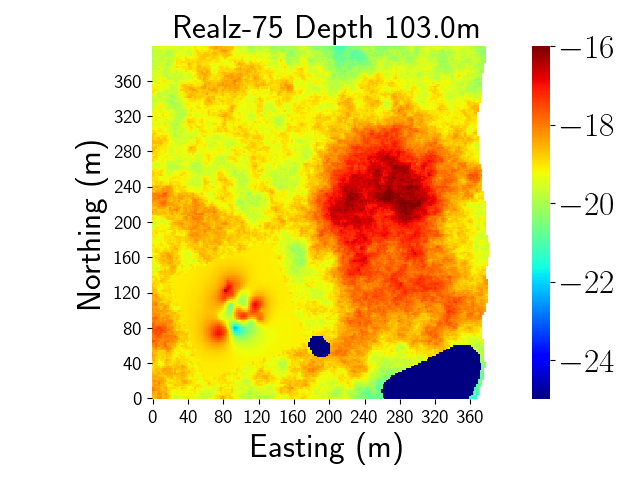}}
  \caption{\textbf{Multiple realizations of 3D permeability field (synthetic data) for hydrogeophysics simulations:}~This figure show samples of the permeability {\color{black}(units are in $\mathrm{m}^2$)} realizations generated for developing hydrogeophysics simulation data.
  The top, middle, and bottom figures show 2D slices of the 3D permeability field {\color{black}(natural log-transformed)} at different depths.
  They also show the spatial patterns of permeability field that DL-enabled inverse models need to capture across training, validation, and testing sets.
  \label{Fig:Perm_Realz_Data}}
\end{figure}

\begin{figure}[!htbp]
  \centering
  \subfigure[Raw data]
    {\includegraphics[width = 0.325\textwidth]
    {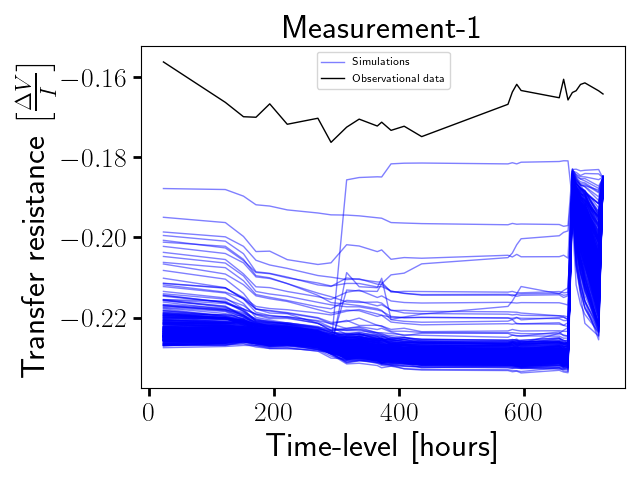}}
  \subfigure[Raw data]
    {\includegraphics[width = 0.325\textwidth]
    {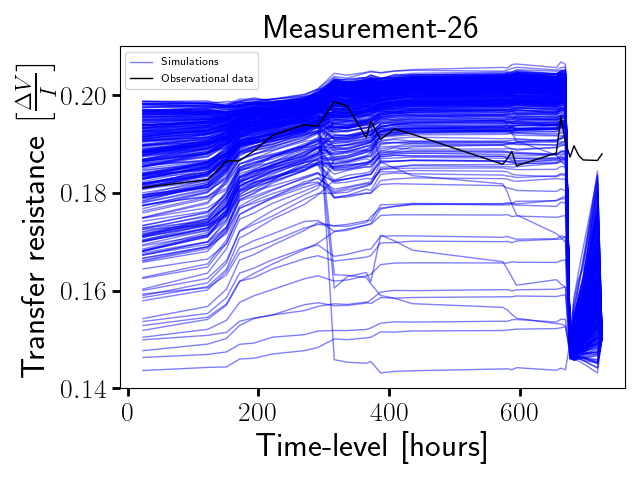}}
  \subfigure[Raw data]
    {\includegraphics[width = 0.325\textwidth]
    {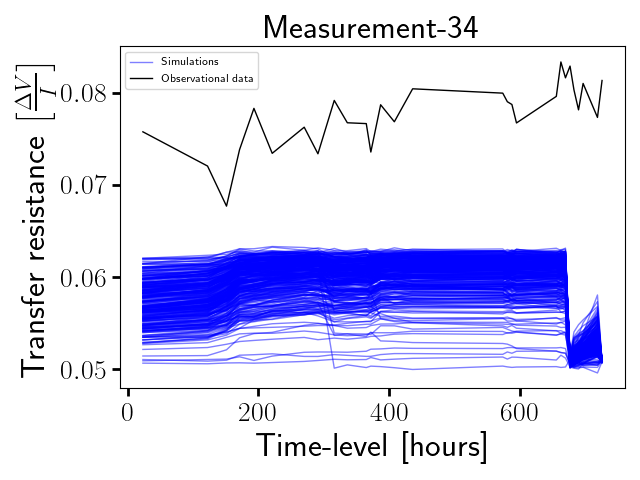}}
  \subfigure[Raw data]
    {\includegraphics[width = 0.325\textwidth]
    {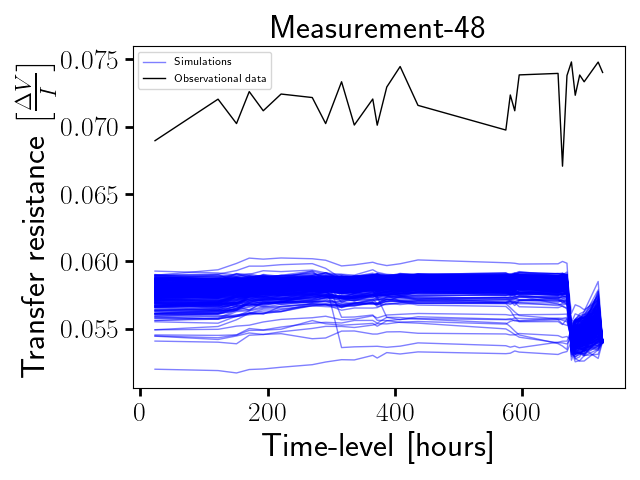}}
  \subfigure[Raw data]
    {\includegraphics[width = 0.325\textwidth]
    {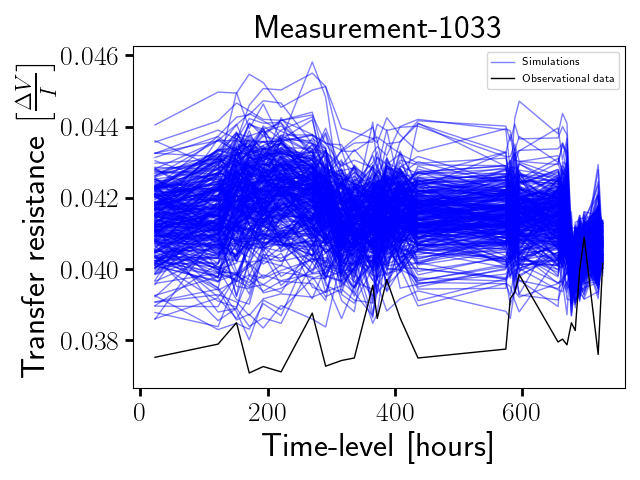}}
  \subfigure[Raw data]
    {\includegraphics[width = 0.325\textwidth]
    {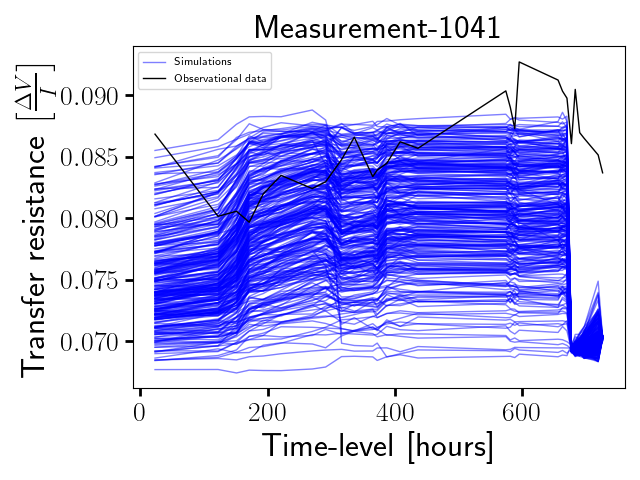}}
  \subfigure[Normalized data]
    {\includegraphics[width = 0.325\textwidth]
    {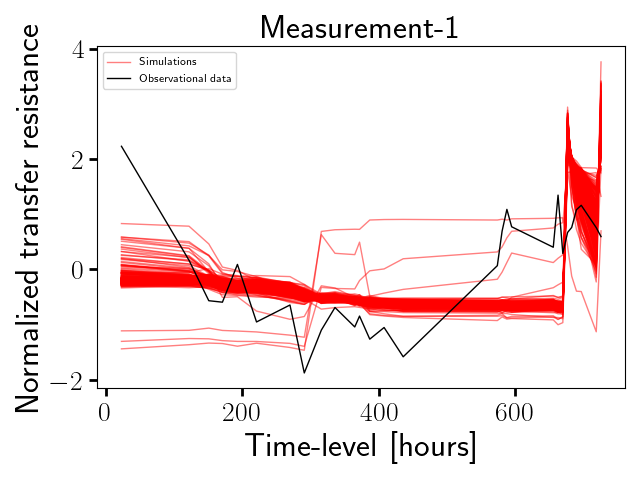}}
  \subfigure[Normalized data]
    {\includegraphics[width = 0.325\textwidth]
    {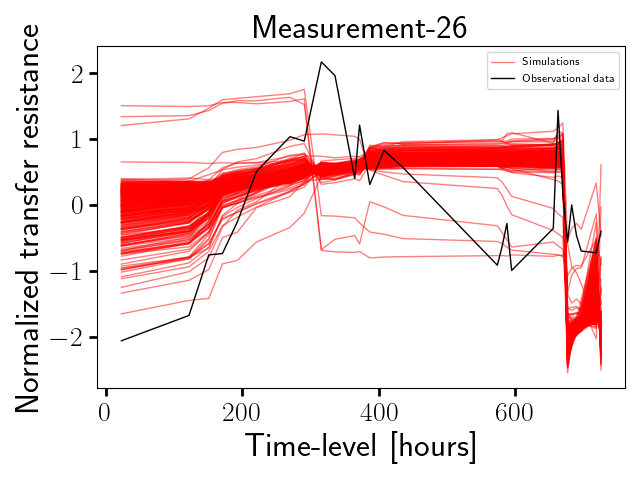}}
  \subfigure[Normalized data]
    {\includegraphics[width = 0.325\textwidth]
    {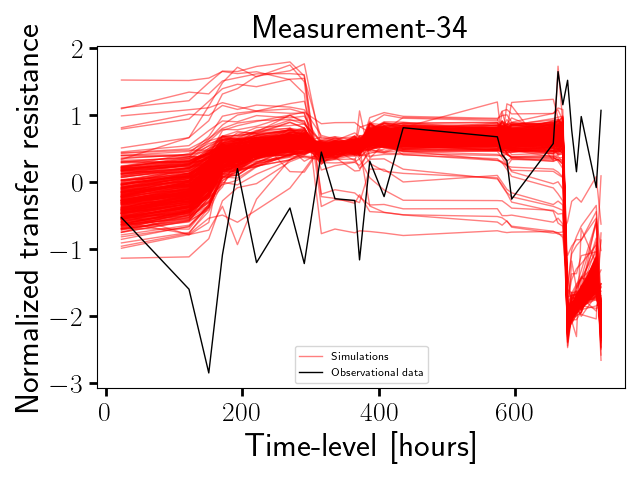}}
  \subfigure[Normalized data]
    {\includegraphics[width = 0.325\textwidth]
    {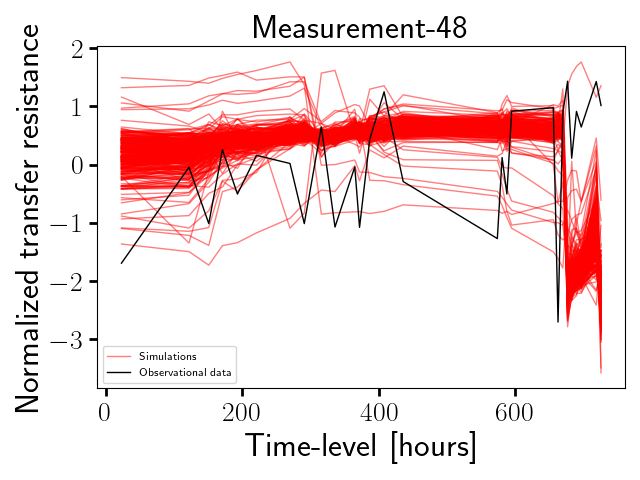}}
    \subfigure[Normalized data]
    {\includegraphics[width = 0.325\textwidth]
    {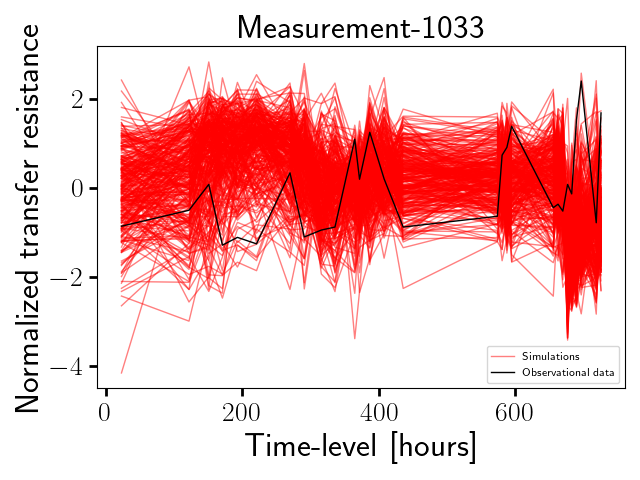}}
  \subfigure[Normalized data]
    {\includegraphics[width = 0.325\textwidth]
    {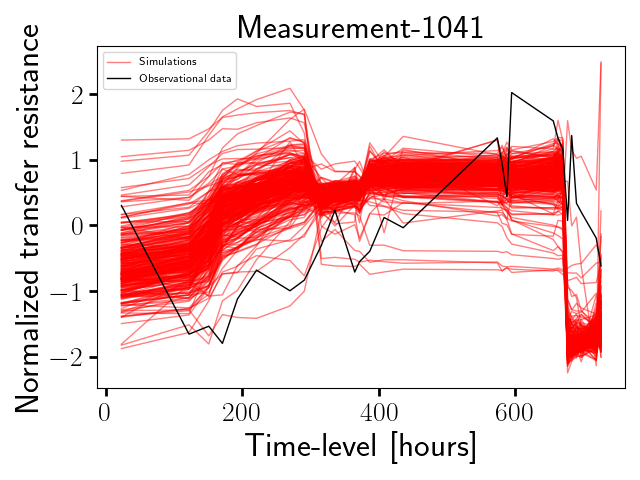}}
  \caption{\textbf{Simulations vs. observational time-lapse ERT data:}~This figure shows samples of raw and normalized transfer resistance among 40,466 measurements.
  Each of the blue and red lines corresponds to raw and normalized simulation data for 28 different times.
  The black line represents the observational data.
  Note that there are 302 blue/red lines for each measurement.  
  From this figure, it is evident that the observational data is contained within the normalized simulation data.
  \label{Fig:ERT_Sim_Obs_Data}}
\end{figure}

High-fidelity numerical simulations are performed to acquire time-lapse ERT surveys for 302 realizations of permeability. 
Figure S1 in supplementary material provides a step-by-step procedure to generate simulation data.
We obtain potential difference measurements at the surface electrodes for a given realization of the 3D permeability field.
These 302 unique pairs of time-lapse ERT and permeability fields are used for training DL-based inverse models.
Figure~\ref{Fig:Perm_Realz_Data} shows multiple realizations of the 3D permeability field at different depths.
Warmer and cooler colors represent high and low permeabilities, respectively.
The values in this figure are shown in $\ln[k(\mathbf{x})]$.
{\color{black}We generated the prior ensemble of the permeability field conditioned on results from Chen et al. (2013) \cite{chen2013application}.}
These results assimilated previous aquifer characterization efforts at the Hanford 300 area, including constant rate injection measurements, flowmeter surveys, and conservative tracer tests.
The posterior realizations of 3D permeability field in a sub-domain of size $120\mathrm{m} \times 120\mathrm{m} \times 15\mathrm{m}$ within the current modeling domain are used to produce realizations of structural parameters.
That is the parameters for the exponential variogram model and conditioning points for generating the hydraulic conductivity field within the \texttt{PFLOTRAN} simulation domain. 
A 3D permeability field is then generated from each realization of structural parameters and conditioning points using the Kriging method \cite{rubin2003applied}. 
A total of 302 realizations of permeability field are included in the prior ensemble.
Table \ref{table:perm_stats} provides the associated statistics of the generated field.
Figure~\ref{Fig:ERT_Sim_Obs_Data} shows the raw and normalized time-lapse ERT simulations.
It compares the generated ensembles (i.e., different transfer resistance measurements) with the observational data.
Each blue/red line represents a specific measurement drawn from the 302 ensembles.
Blackline represents the corresponding field measurement.
A total of 40,466 measurements are assembled from the time-lapse ERT survey.
From this figure, qualitatively, we can see that the trend in the simulated time-lapse ERT is similar to observed data.

\begin{table}[htbp]
    \centering
    \caption{Statistics of 3D permeability field {\color{black}(units are in $\mathrm{m}^2$)} realizations used to perform \texttt{PFLOTRAN} simulations.
    $\mu[\bullet]$ and $\sigma[\bullet]$ are the mean and standard deviation for each realization.}
    \scriptsize
    \begin{tabular}{|c|c|c|}\hline
      \textbf{Permeability -- $k(\mathbf{x})$ {\color{black}[$\mathrm{m}^2$]}} & \textbf{Minimum} & \textbf{Maximum} \\ \hline 
      $\mathrm{min}[k(\mathbf{x})]$ & $8.502 \times 10^{-12}$ & $1.916 \times 10^{-10}$  \\ \hline
      $\mathrm{max}[k(\mathbf{x})]$ & $4.406 \times 10^{-8}$ & $2.502 \times 10^{-6}$  \\ \hline
      $\mathrm{min}[\ln[k(\mathbf{x})]]$ & -25.491 & -22.375 \\ \hline
      $\mathrm{max}[\ln[k(\mathbf{x})]]$ & -16.938 & -12.898  \\ \hline
      $\mu[k(\mathbf{x})]$ & $1.706 \times 10^{-9}$ & $9.351 \times 10^{-8}$  \\ \hline
      $\sigma[k(\mathbf{x})]$ & $1.175 \times 10^{-9}$ & $1.434 \times 10^{-9}$  \\ \hline
      $\mu[\ln[k(\mathbf{x})]]$ & -20.525 & -17.528  \\ \hline
      $\sigma[\ln[k(\mathbf{x})]]$ & 0.412 & 1.793  \\ \hline
    \end{tabular}
    \label{table:perm_stats}
\end{table}

\section{\textbf{PROPOSED SCALABLE DEEP LEARNING METHODOLOGY}}
\label{Sec:S3_DL_Framework}
\begin{figure}[!htbp]
  \centering
    \includegraphics[width = 1.05\textwidth]{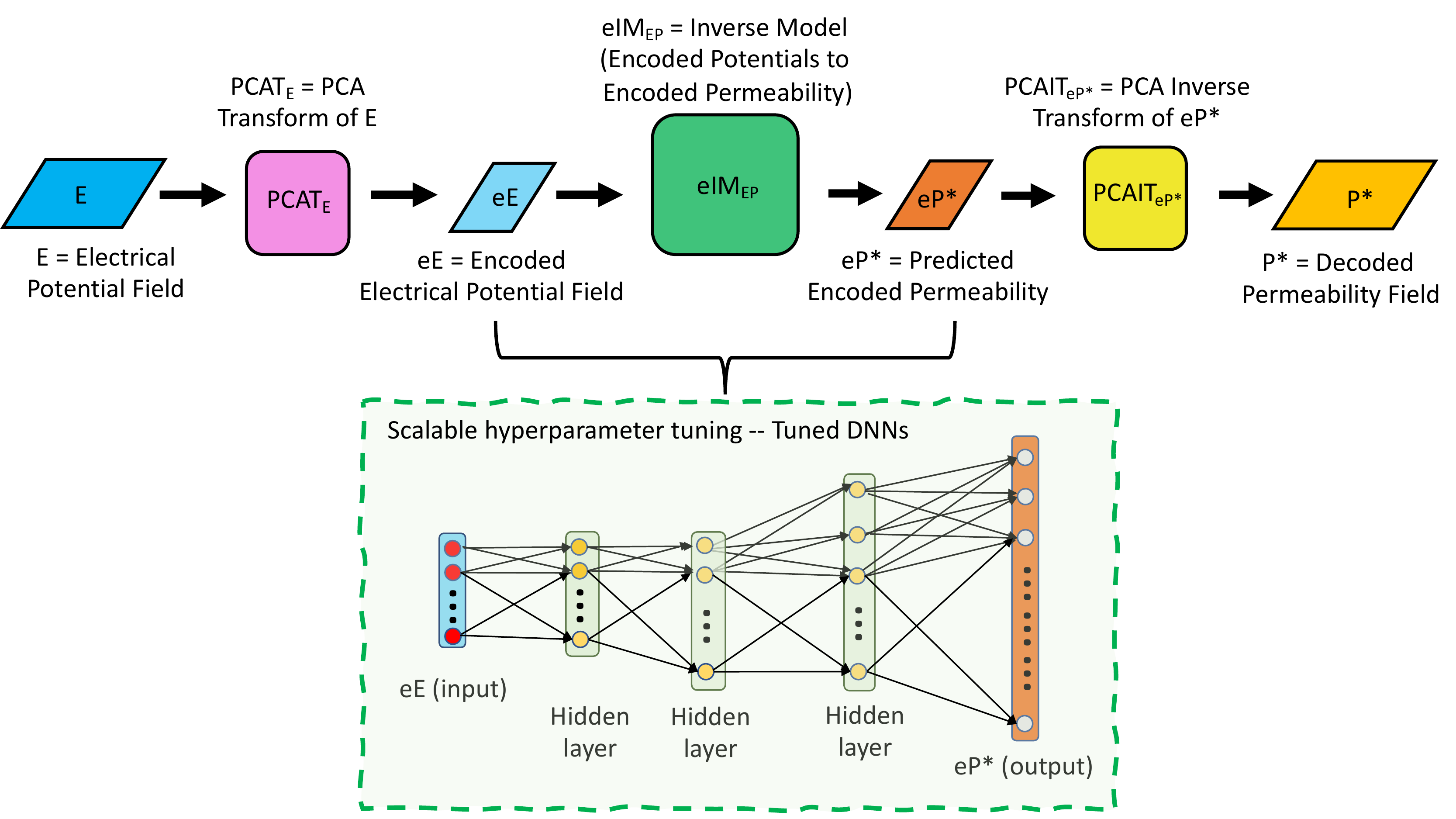}
  \caption{\textbf{Proposed scalable DL methodology:}~A pictorial description of the proposed workflow to develop DL-enabled inverse models.
  The top figure shows the process of mapping the electrical potential field to the permeability field.
  PCA is used to perform dimensionality reduction to encode and decode transfer resistance and 3D permeability.
  The bottom figure shows a fully connected DNN architecture to map the encoded potential to the encoded permeability.
  Hyperparameter tuning (e.g., number of neurons in each layer, learning rate) is performed using high-performance computing resources at NERSC to identify tuned DNN architectures.
  The predicted principal components are decoded using the permeability PCA estimator to obtain the predicted 3D permeability field.
  \label{Fig:DL_Workflow_Architecture}}
\end{figure}

This section presents a methodology based on deep learning to estimate 3D permeability from time-lapse ERT.
Figure~\ref{Fig:DL_Workflow_Architecture} pictorially describes the overall framework to train and tune DL-enabled inverse models.  
{\color{black}The methodology is primarily divided into four main steps:~(1) pre-processing hydrogeophysics simulations, (2) dimensionality reduction using principal component analysis (PCA), (3) mixup-based data augmentation on extracted PCA components, and (4) tuning deep neural network (DNN) models on encoded features at scale.}
We describe each of these steps in the below subsections.

\subsection{\textbf{Data pre-processing}}
\label{SubSec:S3_Data_preprocessing}
The first step in our workflow is data pre-processing.
We pre-process the permeability field and {\color{black} standardize} the time-lapse ERT for subsequent PCA and DNN model training.
{\color{black}Specific cells in the raw 3D permeability field are masked \cite{rubin2010bayesian} before pre-processing.
The masked cells are not used for training PCA and DNN models.
These include river, Ringold, and near-surface cells (between 107m to 110m).}
The unmasked 3D permeability field (a total of 585,453 cells) is flattened and transformed into a natural log scale (e.g., see Fig.~\ref{Fig:Perm_Realz_Data}).
{\color{black}Each 40,466 time-lapse ERT measurement is standardized by calculating the mean and standard deviation across the 28 time-steps.
Then, we remove the mean from the time-series for each realization and then scale it to unit variance.
This standard scaling procedure on each measurement results in a zero mean and unit standard deviation across all the realizations.}
Similarly, the observational data is standardized for each measurement by removing the mean and normalizing it to produce unit standard deviation across all the time-stamps.
Figures~\ref{Fig:ERT_Sim_Obs_Data}(g)--(l)) show samples of normalized by measurements.
Each red line represents a normalized simulation, while the black line represents the observational data.
This normalization by measurement resulted in observations contained within the simulation data.
The pre-processed permeability field and normalized ERT surveys are split into training, validation, and testing sets, which are 282, 10, and 10 simulations.

\subsection{\textbf{PCA-based dimensionality reduction}}
\label{SubSec:S3_PCA_DimRed}
The second step in our workflow is dimensionality reduction.
PCA is a dimensionality reduction method \cite{bro2014principal,ringner2008principal} that uses the singular value decomposition to project data to a lower-dimensional space.
Using PCA on high-dimensional data produces reduced components, capturing and retaining the most crucial information from the original data.
The components or principal axes represent the directions of maximum variance in the input data.
This reduction in input features is a significant advantage as the dimensions of our hydrogeophysics data are extremely large.
Specifically, the time-lapse ERT and 3D permeability feature space dimensions are 1,133,048 and 585,453, respectively. 
We note that developing a deep or convolutional neural network that maps the high-dimensional time-lapse ERT data to the 3D permeability field is computationally intractable.
This is because of the massive number of trainable weights and associated data requirements.
As a result, PCA is used to significantly reduce the dimensions and encode the inputs into components while retaining the most informative aspects of the data.
The permeability and potential field PCA models developed on the training data are then used to transform validation and testing sets.

\begin{figure}[!htbp]
  \centering
    \subfigure[Potential field:~EVR vs. principal components]
    {\includegraphics[width = 0.475\textwidth]{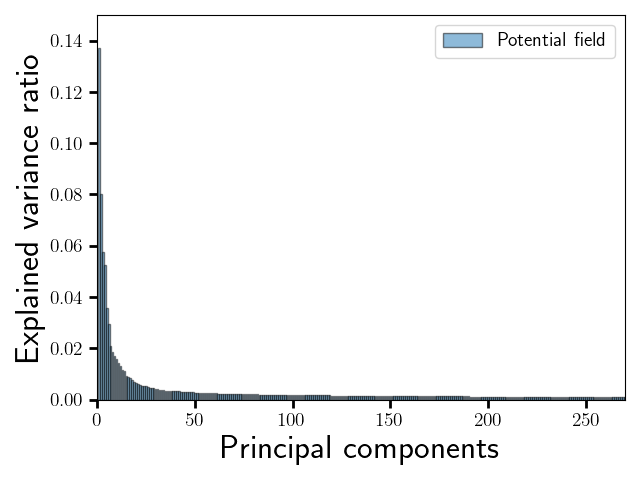}}
    \subfigure[Permeability field:~EVR vs. principal components]
    {\includegraphics[width = 0.475\textwidth]{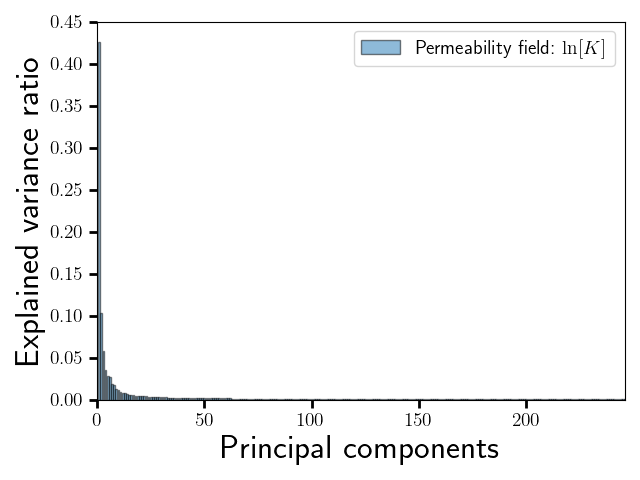}}
  \caption{\textbf{PCA-based dimensionality reduction:}~This figure shows the distribution of variance ratio for all the principal components determined using the PCA estimators of permeability and electrical potential fields.
  The left and right figures show the variances for time-lapse ERT and 3D permeability field, respectively. 
  The PCA estimators are developed on the 282 realizations of training data with 0.99 variances.
  This PCA training resulted in 270 and 246 principal components that represent potential and permeability fields, respectively.
  Each bar in these figures represents a principal component, which explains the variance in the simulation data.
  The extracted components are then used to transform the validation and test datasets.
  \label{Fig:PCA_Variance_eEeP}}
\end{figure}

In this study, we use the \textsf{scikit-learn}'s PCA to fit and transform the data \cite{scikit-learn}.
The permeability and potential PCA models are trained to retain 99\% of the data variance (information).
{\color{black}We have trained PCA estimators using different percentages of data variances (e.g., 80\%, 85\%, 90\%, and 99\%). 
We found that PCA estimator trained using 99\% variance is able to perform better on test datasets.} 
This training process reduces the dimensions of the normalized potential field from 1,133,048 {\color{black}($28 \, \mathrm{time stamps} \times 40,466 \, \mathrm{measurements}$)} to 270 PCA components.
Similarly, the dimensions of pre-processed permeability field are reduced from 585,453 to 246 PCA components.
Figure~\ref{Fig:PCA_Variance_eEeP} shows the explained variance ratio (EVR) of the extracted principal components.
EVR provides the amount of variance explained by each of the principal components.
The top potential and permeability PCA components explain approximately 14\% and 43\% variance in the data, respectively.
The top 20 components explain more than 80\% of variance in ERT and permeability training data.
The remaining 250 (out of 270) and 226 (out of 246) potential and permeability PCA components explain the rest of the 99\% of variance.

\subsection{\textbf{Data augmentation based on mixup}}
\label{SubSec:S3_Data_augmentation}
The third step in our workflow is data augmentation.
We perform mixup \cite{zhang2017mixup} on the extracted PCA components to create additional training samples.
Mixup is a recent data augmentation technique widely used in computer vision to train DL models to temper their prediction overconfidence.
Figure S2 in the supplementary material summarizes our data augmentation strategy to generate new training samples.
This data augmentation technique allows us to develop weakly labeled training samples by convexly combining inputs (PCA components of ERT) and outputs (PCA components of permeability) as follows:
\begin{subequations}
  \label{Eqn:Mixup_and_Variants}
  \begin{align}
    \label{Mixup_x}
    \mathbb{x}_{\textrm{\tiny{mixup}}} &= \lambda
    \mathbb{x}_i + (1 - \lambda) \mathbb{x}_j \\
    \label{Mixup_y}
    \mathbb{y}_{\textrm{\tiny{mixup}}} &= \lambda
    \mathbb{y}_i + (1 - \lambda) \mathbb{y}_j
  \end{align}
\end{subequations}
where $\mathbb{x}_i$ and $\mathbb{x}_j$ are randomly sampled inputs from training set.
$\mathbb{y}_i$ and $\mathbb{y}_j$ are the corresponding training outputs.
$\mathbb{x}_{\textrm{\tiny{mixup}}}$ and $\mathbb{y}_{\textrm{\tiny{mixup}}}$ are the mixed-up samples obtained from the training pairs.
$\lambda \in [0, 1]$ is a random number that determines the mixing ratio.
This number is drawn from a symmetric Beta distribution, $Beta(\alpha,\alpha)$.
$\alpha$ is a parameter that controls the strength of the combination between pairs of inputs and the associated training labels. 

Equations~\ref{Mixup_x}--\ref{Mixup_y} use linear combinations of hidden representation of samples. 
Verma et al., \cite{verma2018manifold} have shown that interpolation of hidden representation can improve the performance of DL models.
A total of 39,621 unique pairs (i.e., $\Comb{282}{2} =\frac{282!}{280! \, 2!}$) are obtained to develop pre-computed mixup samples.
$\alpha$ is assumed to be equal to 0.5.
Based on this value of $\alpha$, ten different $\lambda$ values are generated to combine each strongly labeled training pair.
As a result, 396,210 unique weakly labeled samples are obtained to train DL-enabled inverse models.
Note that the strongly labeled data used to generate mixup samples are not used in this training process.
However, they are used to comparing DL model predictions trained with and without mixup.

\subsection{\textbf{Scalable hyperparameter search and tuned DNN architectures}}
\label{SubSec:S3_tuning}
Our workflow's fourth and final step is to search for optimal DNN architectural parameters (called hyperparameters).
The tuned DNNs map encoded ERT to encoded permeability.
The predicted principal components are then decoded using the trained permeability PCA estimator to obtain a spatially distributed field (see Fig.~\ref{Fig:DL_Workflow_Architecture})

Hyperparameters are different parameters used for training the DNN models. 
They improve the performance of a DNN, resulting in an optimized/tuned architecture.
Hyperparameters explored and tuned in this study are shown in Table~\ref{table:Hyperparameter_yesno_mixup}.
We use \textsf{Keras} API in \textsf{Tensorflow} package \cite{tf-keras} to build our DL-enabled inverse models.
Hyperparameter tuning is the process of searching for the best possible values for the hyperparameters (e.g., number of hidden layers, learning rate) of a DL model.
Tuning the DNN's hyperparameters is essential because these parameters directly control the predictive performance (e.g., accuracy on unseen data) of trained DL-enabled inverse models. 
Hyperparameter tuning requires a search space, which can be significant even for small DNNs. 
The tuning process will become computationally expensive with the increment of the number of hyperparameters. 
The \textsf{Keras Tuner} \cite{omalley2019kerastuner} when combined with \textsf{mpi4py} \cite{dalcin2021mpi4py} or \textsf{multiprocessing} \cite{palach2014parallel} helps to overcome such a challenge\footnotetext[1]{The search for hyperparameters is performed in parallel on NERSC \cite{NERSC2021}, a high performance computing user facility operated by Lawrence Berkeley National Laboratory for the United States Department of Energy Office of Science. Scalable hyperparameter search is achieved by using Python's built-in package called \textsf{multiprocessing} and \textsf{mpi4py} (MPI for Python) package. A total of 136 cores of 1.40 GHz Intel Xeon Phi Processor 7250 with 192 GB DDR4 and 32 GB MCDRAM are used for our hyperparameter tuning. The computational cost to train 1200 DNNs, discover optimal architectures, and post-process the DNN results correspond to approximately 222,000 core hours on NERSC.}.
This library used in this study allows us to pick the optimal set of hyperparameters for DNNs.
It will look for specific points in the search space where the hyperparameter tuning will show its impact on improving model performance.
The \textsf{Keras Tuner}-enabled search process will produce accelerated accuracy on the validation/test datasets, reduce the number of DL model parameters (e.g., number of hidden layers), thereby resulting in faster inference speed (e.g., we can see this from computational cost Section~\ref{SubSec:S4_Comp_Cost}).

Hyperparameter search and tuning are influenced by the random initialization of trainable weights (i.e., the random seed used) in DNN.
Even with fixed hyperparameters, different tuned architectures can be obtained when we perform our tuning.
This study uses 15 popular random seeds \cite{rand_seed_popular_1,rand_seed_popular_2,rand_seed_popular_3} and 40 different architectures trained in each random seed to obtain optimal values.
The optimal hyperparameter set is chosen based on the validation mean squared error using the Bayesian optimization method in the \textsf{Keras Tuner} \cite{omalley2019kerastuner} library.
Supplementary text S3 provides more details on the process of utilizing the Bayesian optimization method to perform hyperparameter tuning for our study.
Table~\ref{table:TunedModels_yesno_mixup} shows the best candidates identified in each of the 15 different random seeds with and without mixup training.
This table shows that the tuning process produced light-weighted DNN architectures (i.e., fewer trainable weights).
In Sec.~\ref{Sec:S4_Results} and supplementary figures, we show the predictions of these 15 different best DL-enabled inverse models and associated errors on training, validation, and test data.

\begin{table}[htbp]
    \centering
    \caption{This table provides the hyperparameter space to develop tuned DNNs for each of 15 different popular random seeds.
    Forty different architectures are trained in each random seed using the \textsf{Keras Tuner} based on the Bayesian Optimization method.
    Batch sizes of 10 and 100 are used to perform training with and without mixup, respectively.
    Epoch sizes of 500 and 4000 are used to train DNNs with and without mixup, respectively.}
    \scriptsize
    \begin{tabular}{|c|c|c|c|}\hline
        \multirow{2}{*}{\textbf{Hyperparameter type}} &
        \multirow{2}{*}{\textbf{Description}} & 
        \multicolumn{2}{c|}{\textbf{Explored options}} \\ \cline{3-4}
        &  & \textbf{No-mixup} & \textbf{Mixup} \\ \hline
        Learning rate & The value of the optimizer in the Adam algorithm & $10^{-6} \times [1, 2, 4, 6, 8, 10,$ &  $10^{-7} \times [1, 5, 10, 20, 40, 60, 80$ \\
        &  & $20, 40, 60, 80, 100, 10^3]$ &  $100, 200, 400, 600, 800, 10^3, 10^4]$ \\ \hline
        Layers & Number of hidden layers & \multicolumn{2}{c|}{$[1, 2, 3, 4, 5]$} \\ \hline
        Nodes & The number of nodes in a hidden layer & \multicolumn{2}{c|}{mininum = 270, maximum = 300, step size = 10} \\ \hline
        Alpha & Slope coefficient in LeakyReLU activation function & \multicolumn{2}{c|}{mininum = 0.0, maximum = 0.5, step size = 0.05} \\ \hline
        Dropout rate & Applies dropout to the input\footnotemark & \multicolumn{2}{c|}{mininum = 0.0, maximum = 0.5, step size = 0.05} \\ \hline
    \end{tabular}
    \label{table:Hyperparameter_yesno_mixup}
\end{table}
\footnotetext[2]{Table-\ref{table:Hyperparameter_yesno_mixup}:~To reduce model over-fitting, we randomly set hidden layer units that connect to the next hidden layer or output to 0 at each step during training time.
The rate value controls the frequency of dropping the units.}

\begin{table}[htbp]
    \centering
    \caption{This table provides the specifics of tuned DNN architectures (among 600 trained models) for 15 different popular random seeds.}
    \scriptsize
    \begin{tabular}{|c|c|c|c|c|c|c|c|c|c|c|}\hline
        \multirow{2}{*}{\textbf{Random seed}} &
        \multicolumn{2}{c|}{\textbf{No. of hidden layers}} & 
        \multicolumn{2}{c|}{\textbf{No. of nodes}} & 
        \multicolumn{2}{c|}{\textbf{Alpha value}} & 
        \multicolumn{2}{c|}{\textbf{Dropout rate}} & 
        \multicolumn{2}{c|}{\textbf{Learning rate}} \\ \cline{2-11}
        & \textbf{No-mixup} & \textbf{mixup} & \textbf{No-mixup} & \textbf{mixup} & \textbf{No-mixup} & \textbf{mixup} & \textbf{No-mixup} & \textbf{mixup} & \textbf{No-mixup} & \textbf{mixup} \\ \hline
        1337 & \multicolumn{2}{c|}{1} & 300 & 270 & 0.5 & 0.15 & \multicolumn{2}{c|}{0.5} & \multicolumn{2}{c|}{$10^{-3}$} \\ \hline
        0 & \multicolumn{2}{c|}{1} & 270 & 300 & \multicolumn{2}{c|}{0.5} & \multicolumn{2}{c|}{0.5} & \multicolumn{2}{c|}{$10^{-3}$} \\ \hline
        1 & \multicolumn{2}{c|}{1} & \multicolumn{2}{c|}{300} & \multicolumn{2}{c|}{0.5} & \multicolumn{2}{c|}{0.5} & \multicolumn{2}{c|}{$10^{-3}$} \\ \hline
        2 & \multicolumn{2}{c|}{1} & \multicolumn{2}{c|}{270} & \multicolumn{2}{c|}{0.5} & \multicolumn{2}{c|}{0.5} & \multicolumn{2}{c|}{$10^{-3}$} \\ \hline
        3 & \multicolumn{2}{c|}{1} & 270 & 300 & 0.45 & 0.5 & \multicolumn{2}{c|}{0.5} & \multicolumn{2}{c|}{$10^{-3}$} \\ \hline
        4 & \multicolumn{2}{c|}{1} & 270 & 300 & \multicolumn{2}{c|}{0.5} & \multicolumn{2}{c|}{0.5} & \multicolumn{2}{c|}{$10^{-3}$} \\ \hline
        5 & \multicolumn{2}{c|}{1} & 270 & 300 & 0.3 & 0.5 & \multicolumn{2}{c|}{0.5} & \multicolumn{2}{c|}{$10^{-3}$} \\ \hline
        7 & \multicolumn{2}{c|}{1} & \multicolumn{2}{c|}{290} & \multicolumn{2}{c|}{0.5} & \multicolumn{2}{c|}{0.5} & \multicolumn{2}{c|}{$10^{-3}$} \\ \hline
        8 & \multicolumn{2}{c|}{1} & \multicolumn{2}{c|}{300} & \multicolumn{2}{c|}{0.5} & \multicolumn{2}{c|}{0.5} & \multicolumn{2}{c|}{$10^{-3}$} \\ \hline
        10 & \multicolumn{2}{c|}{1} & 290 & 300 & \multicolumn{2}{c|}{0.5} & \multicolumn{2}{c|}{0.5} & \multicolumn{2}{c|}{$10^{-3}$} \\ \hline
        11 & 5 & 1 & [300, 270, 270, & 300 & \multicolumn{2}{c|}{0.5} & 0.0 & 0.5 & \multicolumn{2}{c|}{$10^{-3}$} \\
        & & & 300, 270] & & \multicolumn{2}{c|}{} & & & \multicolumn{2}{c|}{} \\ \hline
        13 & \multicolumn{2}{c|}{1} & 290 & 280 & \multicolumn{2}{c|}{0.5} & \multicolumn{2}{c|}{0.5} & \multicolumn{2}{c|}{$10^{-3}$} \\ \hline
        42 & 2 & 1 & [280, 280] & 280 & 0.5 & 0.25 & 0.35 & 0.5 & \multicolumn{2}{c|}{$10^{-3}$} \\ \hline
        100 & 2 & 1 & [280, 300] & 300 & 0.5 & 0.2 & \multicolumn{2}{c|}{0.5} & \multicolumn{2}{c|}{$10^{-3}$} \\ \hline
        113 & \multicolumn{2}{c|}{1} & 270 & 300 & \multicolumn{2}{c|}{0.5} & \multicolumn{2}{c|}{0.5} & \multicolumn{2}{c|}{$10^{-3}$} \\ \hline
    \end{tabular}
    \label{table:TunedModels_yesno_mixup}
\end{table}

\subsection{\textbf{Computational cost}}
\label{SubSec:S4_Comp_Cost}
The simulation data for training DL models are generated using leadership-class supercomputing resources at the NERSC user facility \cite{NERSC2021}. 
The computational cost to run a single \texttt{PFLOTRAN} forward model simulation is approximately 22 minutes on four Cori Haswell compute nodes \cite{NERSC-Cori}.
Each node has two 2.3 GHz 16-core Haswell processors (Intel Xeon Processor E5-2698 v3).
Developing \texttt{PFLOTRAN} simulation data accounts for approximately 14,187 processor hours (i.e, $\approx 0.367 \times 128 \times 302$).
To run a single \texttt{E4D} forward model simulation, it takes approximately 4.5 hours on eight Cori KNL compute nodes.
Each node is a single Intel Xeon Phi Processor 7250 (Knights Landing) with 68 cores $@$ 1.4 GHz.
A total of 2,416 KNL nodes were used to generate \texttt{E4D} simulation data, which is equal to 739,296 processor hours (i.e, $4.5 \times 68 \times (8 \times 302)$).

We have trained the tuned DL-enabled inverse models on a MacBook Pro Laptop (2.3GHz Intel Core i9, 8-core CPU, 64 GB 2667 MHz DDR4 RAM).
The individual computational cost to perform dimensionality reduction using PCA on permeability field and time-lapse ERT is 37 and 25 seconds, respectively.
Training the proposed encoded DNNs for 4000 epochs takes about 60 seconds without mixup.
With mixup-based data augmentation, depending on the DNN architecture, the training times are in the $\mathcal{O}(10^2)-(10^3)$ seconds for a total of 500 epochs.
The total inference wall clock time is approximately 0.19 seconds.
This inference includes encoding a new time-lapse ERT test data using PCA ($\approx 2.0 \times 10^{-2}$ seconds), estimating encoded permeability field using trained DNN ($\approx 8.58 \times 10^{-4}$ seconds), and decoding the estimated permeability using PCA ($\approx 0.167$ seconds).

Based on the above values, it is evident that generating training data is computationally expensive and requires HPC resources.
Total training dataset generation time is in the order of $\mathcal{O}(10^5)$ processor hours.
For instance, it takes a total of $2.5 \times 10^{3}$ processor hours to perform a single forward model simulation of \texttt{PFLOTRAN} and \texttt{E4D}.
This shows that it may be expensive to estimate 3D permeability using traditional inversion or software packages and tools (e.g., \texttt{PEST}, \texttt{DAKOTA}, \texttt{DART}), which require 100s to 1000s of forward models runs.
These multiple forward runs translate anywhere from $\mathcal{O}(10^5)-\mathcal{O}(10^6)$ processor hours. 
Though the computational cost to develop the proposed DL-enabled inverse model is expensive (e.g., primarily training data generation is costly), it is attractive for data assimilation.
For every new time-lapse ERT field study (or ingesting a new realization), traditional data assimilation requires at least $\mathcal{O}(10^3)$ processor hours (e.g., through one or multiple hydrogeophysics forward models runs).
The limited availability and continuous access (e.g., nodes, machine hours, constraints on maximum wall clock time to perform runs) to HPC resources make it challenging to use such state-of-the-art parameter estimation softwares.
Specifically, constraints on the HPC resources may pose a challenge to perform the traditional inversion.
However, once the DL-enabled inverse model is trained, it takes $\mathcal{O}(10^{-2})$ seconds to perform an inversion, thereby resulting in massive computational savings (e.g., a minimum of $\mathcal{O}(10^7)$ seconds).

\section{\textbf{RESULTS AND DISCUSSION}}
\label{Sec:S4_Results}
In this section, we present results to evaluate the overall accuracy of the proposed DL methodology.
First, we will describe the PCA modeling results and associated errors due to encoding and decoding the 3D permeability field.
Second, we present the results for tuned DNN architectures trained using no mixup and pre-computed mixup after PCA.
Various performance metrics are provided to assess trained DL-enabled inverse models (i.e., tuned DNN and PCA estimations).
Finally, we discuss the limitations of the proposed method and possible approaches to overcome such challenges.

\begin{figure}[!htbp]
  \centering
    \subfigure[Training:~Ground truth vs. prediction]
    {\includegraphics[width = 0.32\textwidth]{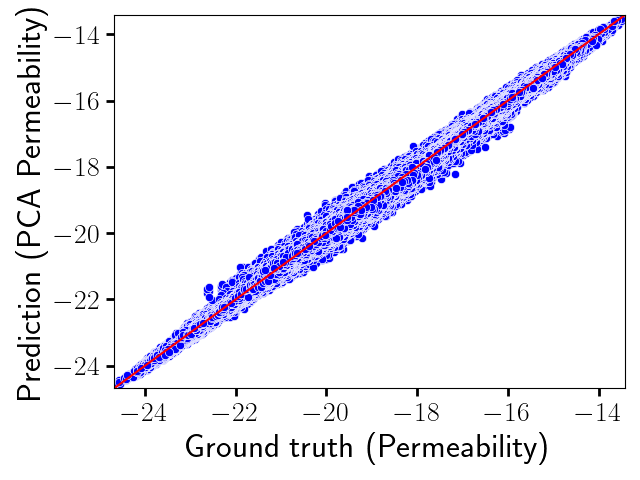}}
    \subfigure[Validation:~Ground truth vs. prediction]
    {\includegraphics[width = 0.32\textwidth]{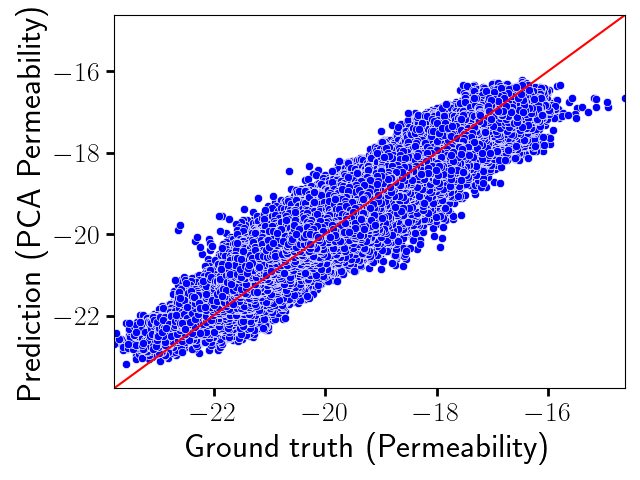}}
    \subfigure[Testing:~Ground truth vs. prediction]
    {\includegraphics[width = 0.32\textwidth]{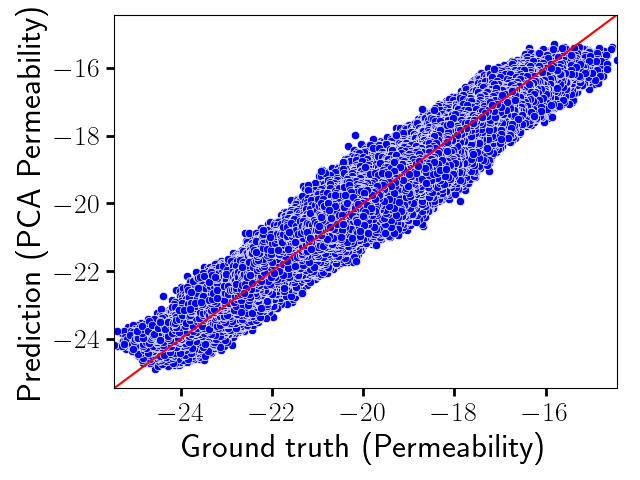}}
    \subfigure[Training:~PCA reconstruction error]
    {\includegraphics[width = 0.325\textwidth]{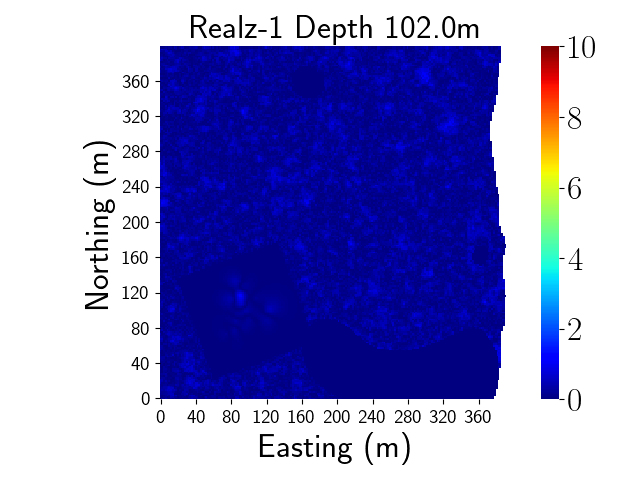}}
    \subfigure[Validation:~PCA reconstruction error]
    {\includegraphics[width = 0.325\textwidth]{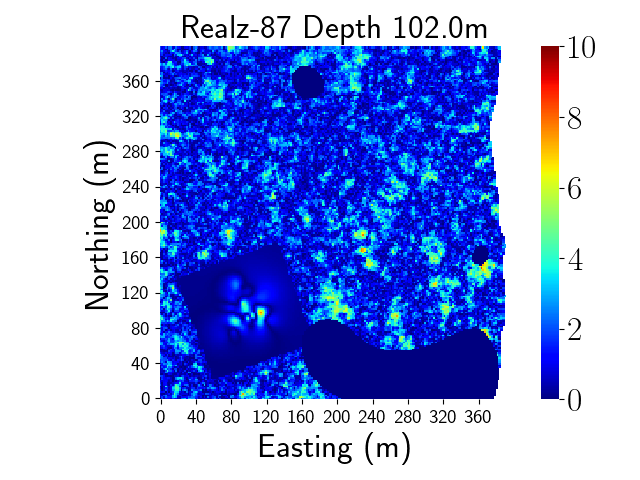}}
    \subfigure[Testing:~PCA reconstruction error]
    {\includegraphics[width = 0.325\textwidth]{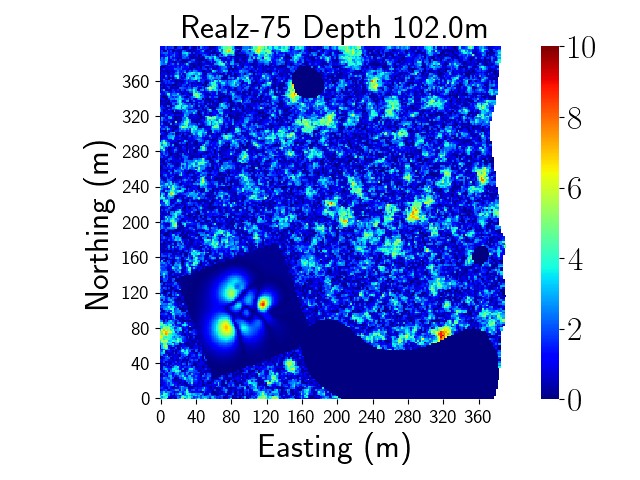}}
    \subfigure[PDF vs. PCA reconstruction error]
    {\includegraphics[width = 0.375\textwidth]{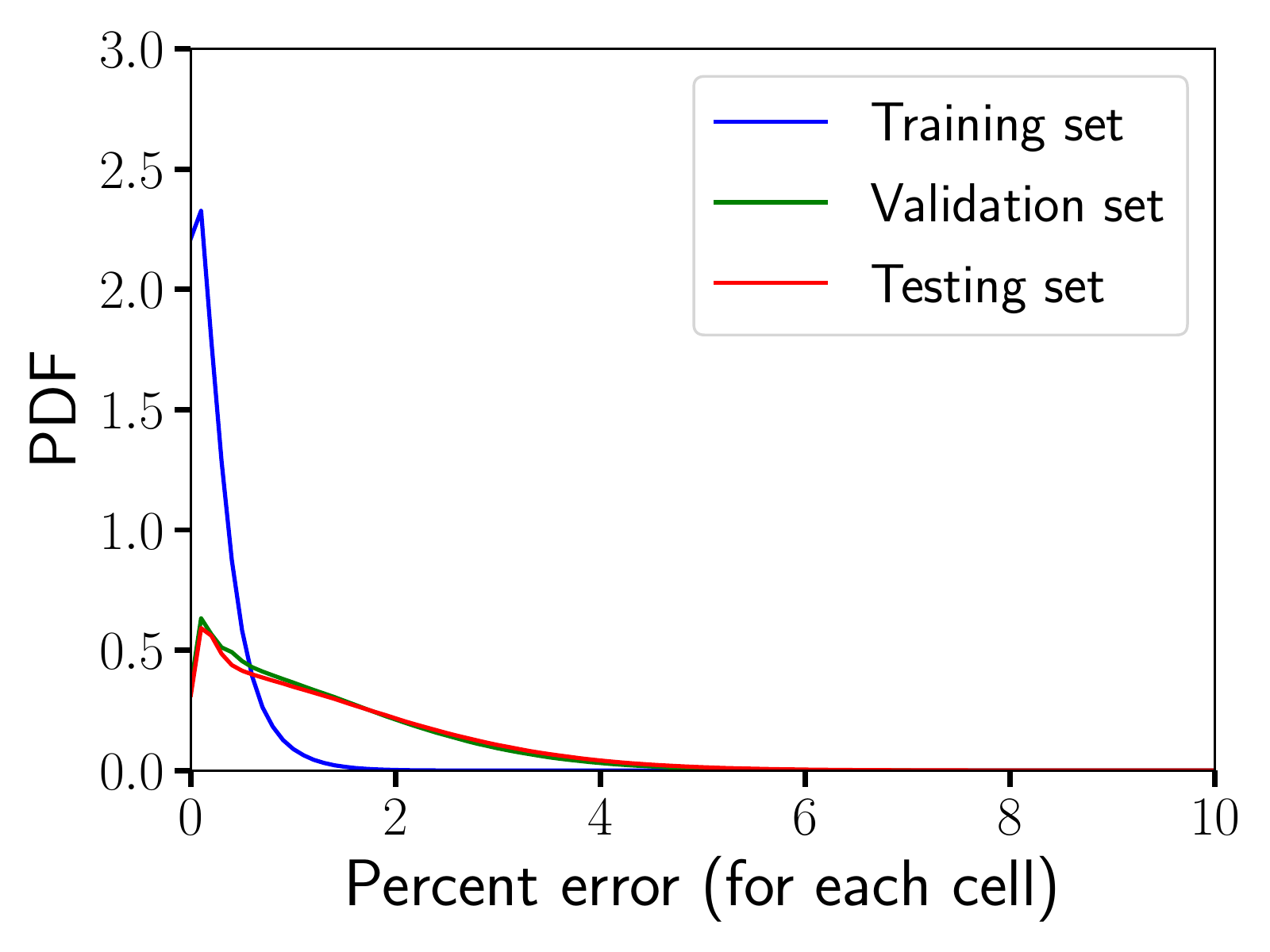}}
  \caption{\textbf{PCA reconstruction and associated percent errors:}~This figure shows the permeability PCA estimator predictions and its errors on training, validation, and test datasets.
  The top figure shows one-to-one plots of PCA estimator predictions (values in $\ln[k(\mathbf{x})]$) with the ground truth for all the realizations in each set.
  To be specific, each blue dot in (a), (b), and (c) represents a permeability value in a 3D grid cell.
  The middle figures show spatial patterns of reconstruction errors (values in percentage) for 2D slices shown in Fig.~\ref{Fig:Perm_Realz_Data}(a), (d), and (g).
  The bottom figure shows the PDF as a function of percent errors in each grid cell for all the training, validation, and test realizations.
  From spatial plots and PDF figure, it is evident that the error in each cell is less than 10\%.
  \label{Fig:PCA_Reconstruction_Errors}}
\end{figure}

Figure~\ref{Fig:PCA_Reconstruction_Errors} provides the predictions and associated percent errors of permeability PCA estimator trained on 282 realizations.
The one-to-one plots in Fig.~\ref{Fig:PCA_Reconstruction_Errors}(a)-(c) compare the ground truth and PCA predictions for all the training, validation, and test data.
These figures show that PCA predictions are scattered near the red line for training data.
However, the predictions are dispersed a little far from the one-to-one line for validation and testing realizations.
For high permeability, we can see considerable under-predictions from PCA estimators, while for lower permeability, we can see over-predictions.
Such a reduced performance of PCA estimators is well-known due to the curse of dimensionality \cite{bro2014principal}.
Very high-dimensional inputs (e.g., we have more than 500,000 permeability features) often make a dimensionality reduction challenging and degrade the PCA encoder-decoder performance.
The spatial distribution of reconstruction errors are shown in Fig.~\ref{Fig:PCA_Reconstruction_Errors}(d)-(f).
The color bar values in these figures are given in percentage.
The associated ground truth for which we calculate the percent errors are shown in Fig.~\ref{Fig:Perm_Realz_Data}(a), (d), and (g).
The spatial patterns for validation and testing realizations reveal that errors are less than 5\% in most parts of the region.
We can also make similar inferences from the probability density function (PDF) plots shown in Fig.~\ref{Fig:PCA_Reconstruction_Errors}(g).
PDFs are computed using kernel density estimation algorithms available in \textsf{statsmodels} python package \cite{seabold2010statsmodels}.
For training realizations, the density is primarily concentrated within the top 2\%.
However, for validation and test sets, the PDF curves have a long tail and spread until 6-8\%.
For all realizations, the percent error in each grid cell is less than 10\% with a probability of 0.99.

\begin{figure}[!htbp]
  \centering
    \subfigure[No mixup]
    {\includegraphics[width = 0.465\textwidth]{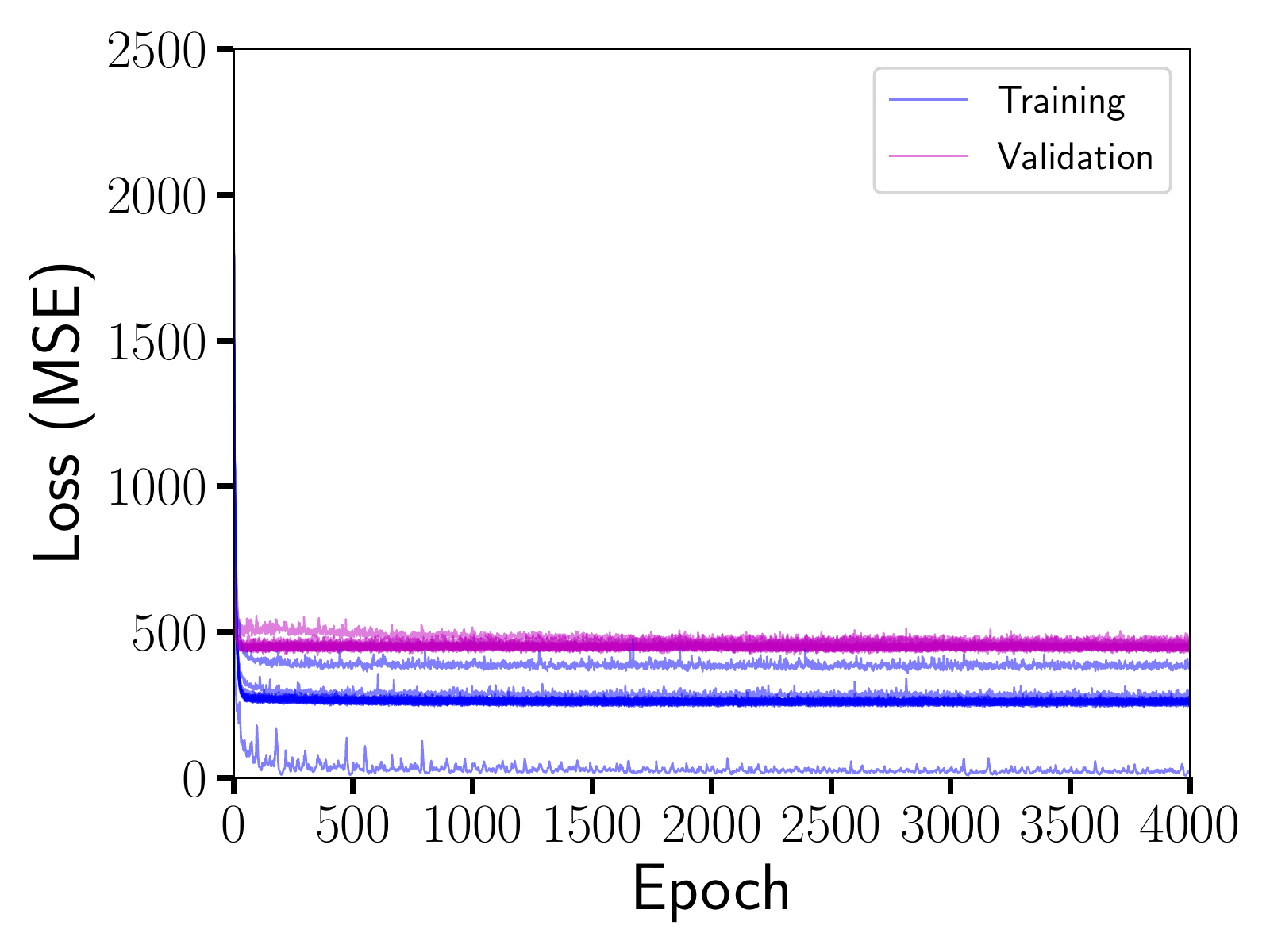}}
    \subfigure[Pre-computed mixup after PCA]
    {\includegraphics[width = 0.465\textwidth]{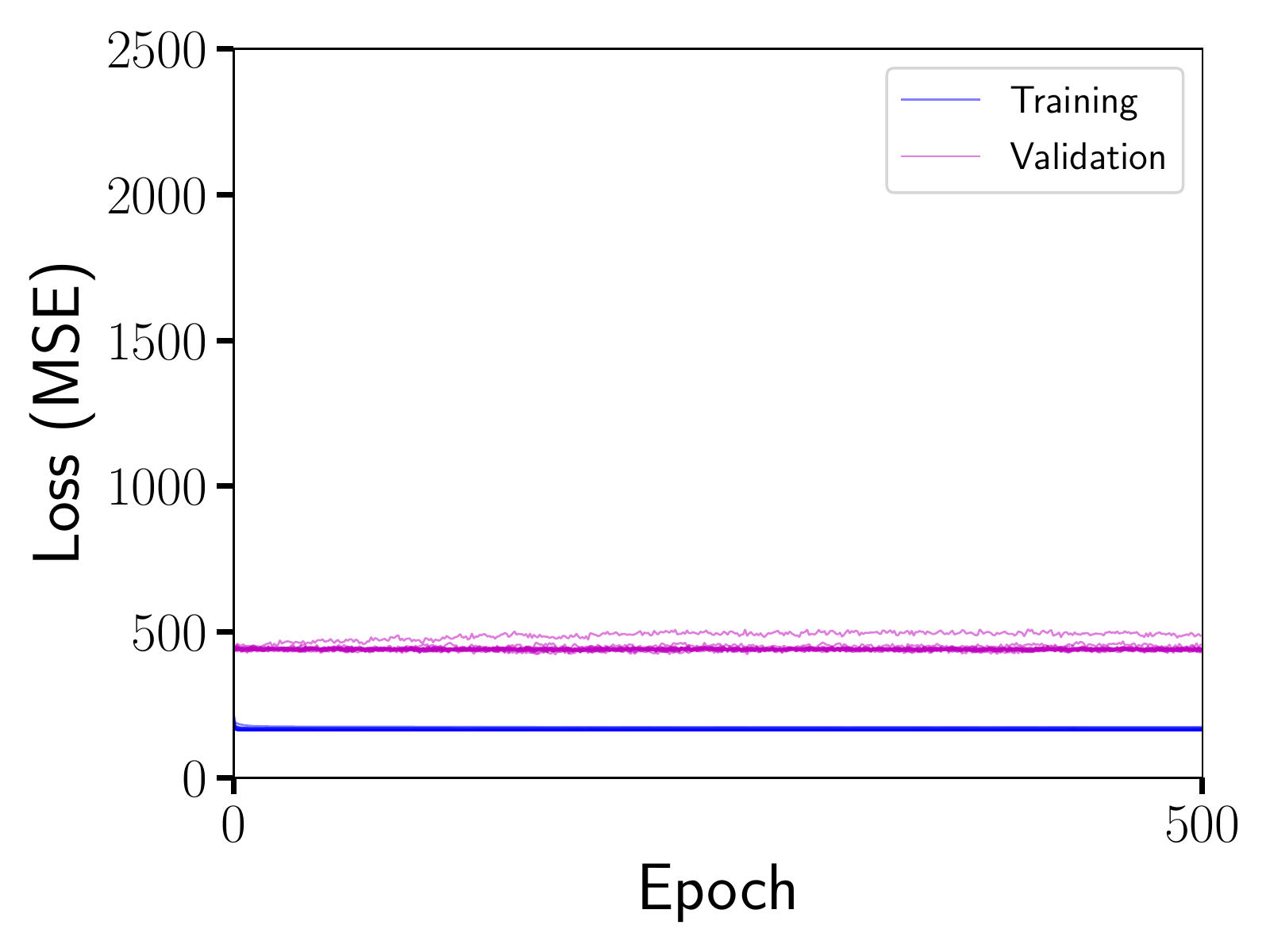}}
  \caption{\textbf{Loss metrics for 15 different best inverse models:}~This figure shows the overall training and validation loss of tuned DL-enabled inverse models for 15 different random seeds.
  The DNN architectures are tuned using strong (no mixup) and weak labels (without mixup).
  Based on the validation loss, it is evident that mixup training has slightly better generalization than no mixup.
  This shows that weakly labeled data slightly helps to reduce over-fitting.
  However, we can see that the DL-enabled inverse models still over-fits the non-mixup and mixup data.
  \label{Fig:Loss_Function_No_Precomp_Mixup}}
\end{figure}

Figure~\ref{Fig:Loss_Function_No_Precomp_Mixup} shows the training and validation loss of 15 different tuned DNNs to estimate the encoded permeability.
The supplementary text shows the individual loss plots of each DNN (trained with and without mixup) tuned using a different random seed in Fig.~S3-S10. 
The training and validation losses are based on mean squared error.
From these figures, in most scenarios, it is clear that training MSE flatten out early, and validation loss slightly increases, which is a sign of over-fitting.
This increase in loss usually happens when DNNs try to capture less informative principal components (i.e., for PCA components that have very low explained variance ratios).
Mixing up PCA components helps to reduce the MSE value (with a lesser number of epochs), but over-fitting still occurs.

\begin{figure}[!htbp]
  \centering
    \subfigure[No mixup:~Training]
    {\includegraphics[width = 0.31\textwidth]{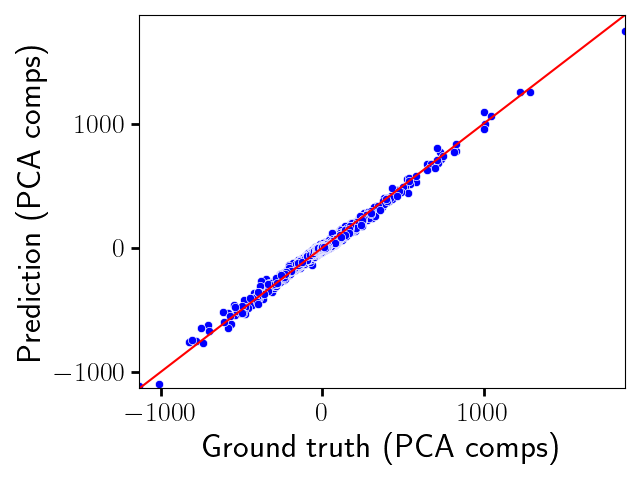}}
    \subfigure[No mixup:~Validation]
    {\includegraphics[width = 0.31\textwidth]{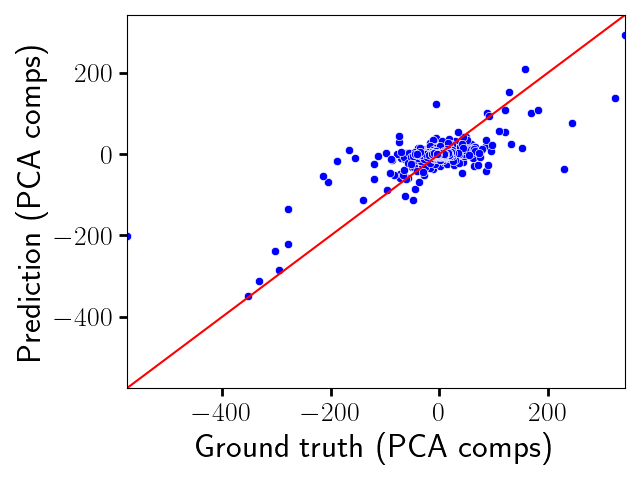}}
    \subfigure[No mixup:~Testing]
    {\includegraphics[width = 0.31\textwidth]{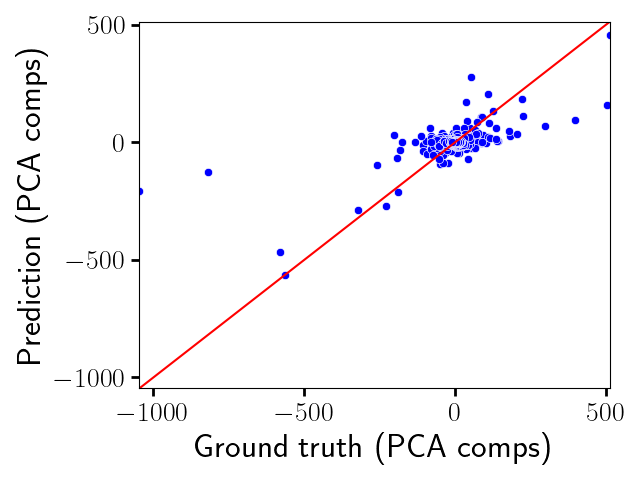}}
    \subfigure[Mixup:~Training]{\includegraphics[width = 0.31\textwidth]{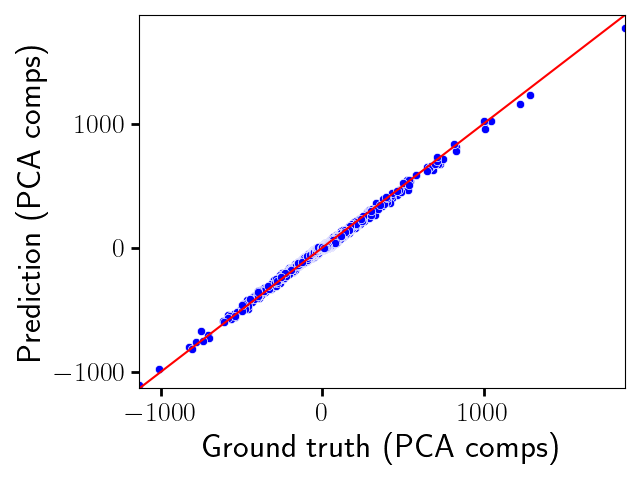}}
    \subfigure[Mixup:~Validation]
    {\includegraphics[width = 0.31\textwidth]{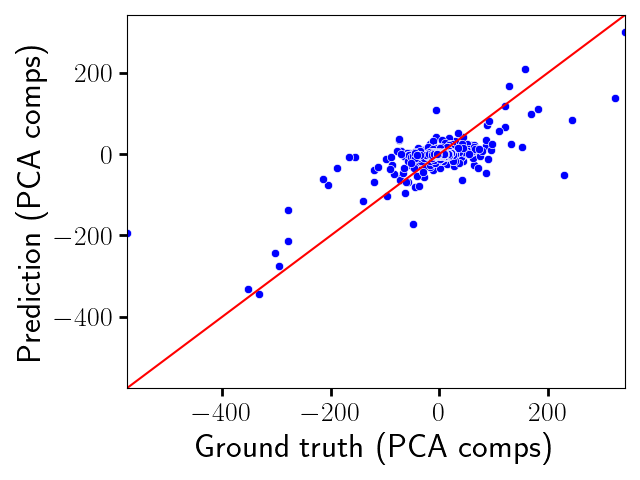}}
    \subfigure[Mixup:~Testing]
    {\includegraphics[width = 0.31\textwidth]{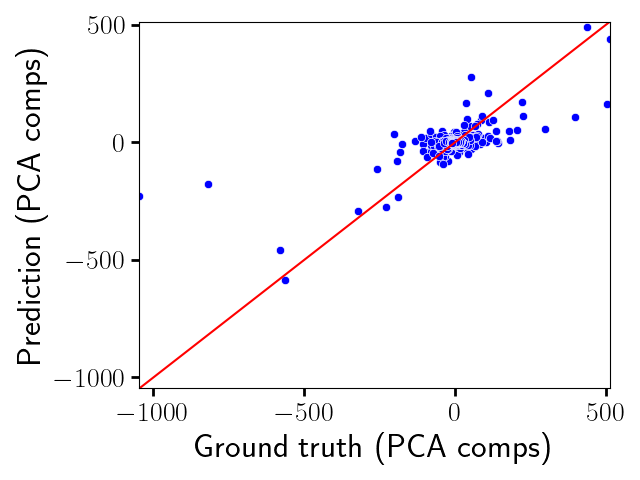}}
    \subfigure[No mixup:~Training]
    {\includegraphics[width = 0.31\textwidth]{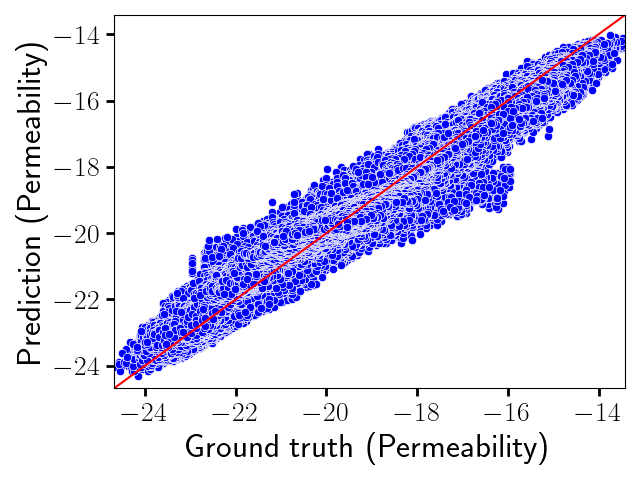}}
    \subfigure[No mixup:~Validation]
    {\includegraphics[width = 0.31\textwidth]{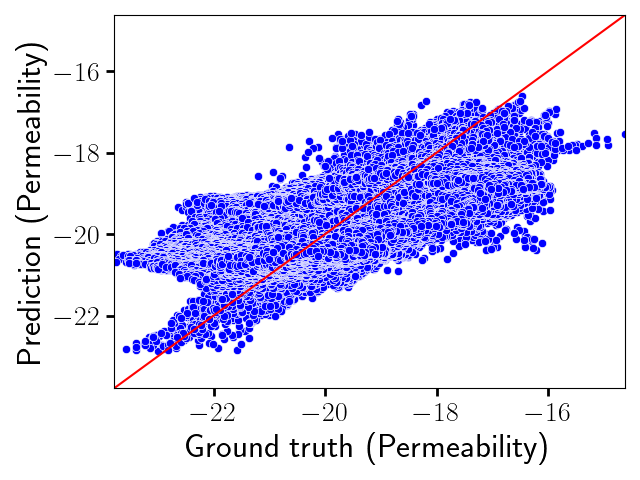}}
    \subfigure[No mixup:~Testing]
    {\includegraphics[width = 0.31\textwidth]{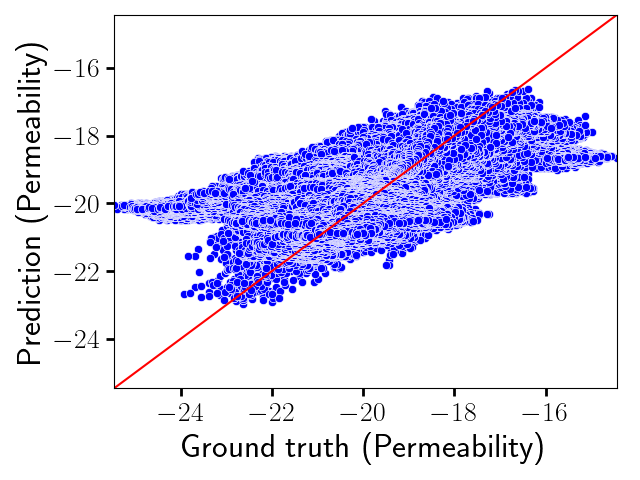}}
    \subfigure[Mixup:~Training]
    {\includegraphics[width = 0.31\textwidth]{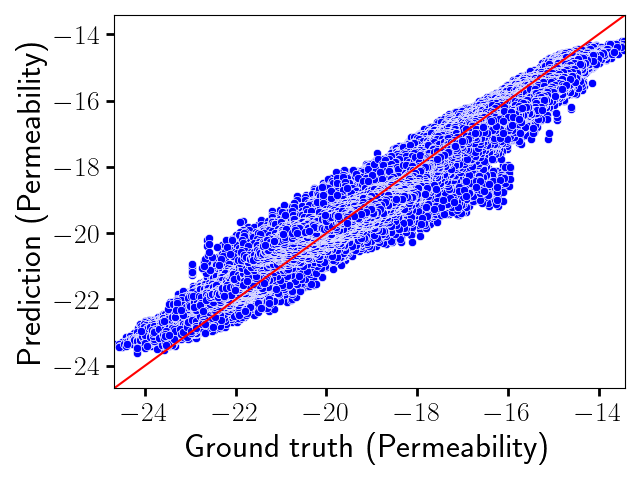}}
    \subfigure[Mixup:~Validation]
    {\includegraphics[width = 0.31\textwidth]{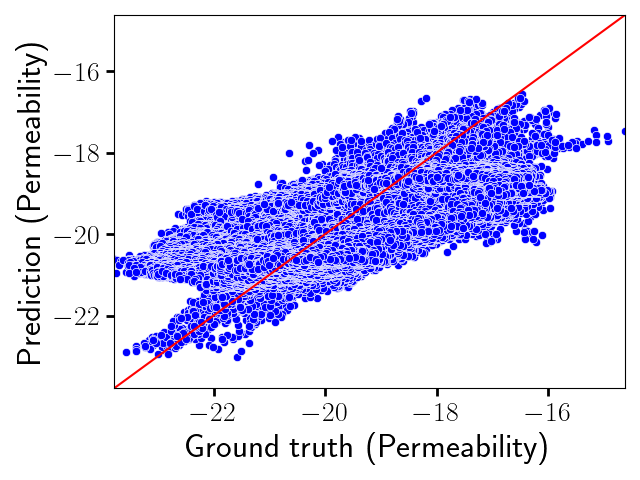}}
    \subfigure[Mixup:~Testing]
    {\includegraphics[width = 0.31\textwidth]{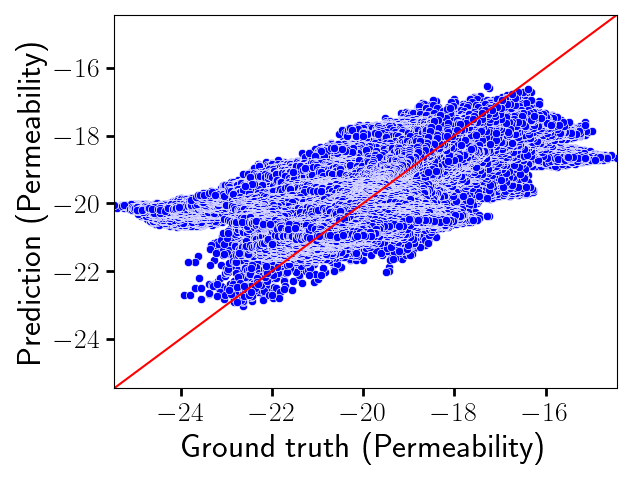}}
  \caption{\textbf{Ground truth vs. Prediction (PCA components and permeability):}~This figure shows one-to-one plots of a tuned DL-enabled inverse model (based on random seed = 0) for encoded and decoded 3D permeability.
  The top and middle figures show the PCA components predictions based on the tuned DNN architecture with and without mixup.
  Each blue dot represents an encoded permeability field.
  The lower set of figures show the predictions of decoded permeability field with and without mixup.
  These predictions are obtained by utilizing the trained PCA estimator on the DL predicted principal components.
  Each blue dot represents a permeability value in a grid cell.
  The permeability predictions based on DL-enabled inverse models trained using mixup samples seem to have slightly higher permeability values than no mixup.
  \label{Fig:True_vs_Pred_PermandPCA}}
\end{figure}

Figure~\ref{Fig:True_vs_Pred_PermandPCA} compares the predictions of DL-enabled inverse models against ground truth.
The scatter plots shown in Figs.~\ref{Fig:True_vs_Pred_PermandPCA}(a)-(c) show the PCA component predictions of DNNs trained without mixup while Figs.~\ref{Fig:True_vs_Pred_PermandPCA}(d)-(f) provide model predictions with mixup.
The trained PCA estimator decodes the DNN predictions to obtain permeability values in each grid cell.
The associated scattered plots for the estimated permeability field are shown in Figs.~\ref{Fig:True_vs_Pred_PermandPCA}(g)-(l).
The results show in Fig.~\ref{Fig:True_vs_Pred_PermandPCA} are based on a tuned DNN (single hidden layers with either 270 or 300 nodes, see Table.~\ref{table:TunedModels_yesno_mixup}) with random seed = 0.  
The figures S3-S18 in the supplementary text provide the predictions and compare them with ground truth for all the remaining tuned DNNs. 
The training predictions show that PCA components are close to the one-to-one line.
The permeability predictions in each grid cell for training data show that mixup performs slightly better than no mixup, as evident from the spread of the blue dots.
However, the DNN predictions of PCA components for both validation and test sets are dispersed due to over-fitting.
From the scatter plots of encoded DNN predictions, it is clear that specific high and low-value PCA components (e.g., values between -1000 and -500, values closer to 500) are estimated with reduced accuracy.
These skewed estimations are propagated when decoding the predicted components using PCA.
As a result, we can see the increased deviation of permeability predictions from the red line.
On validation and test realizations, we see under and over predictions for high and low values of permeabilities.
However, for test realizations, we can see that mixup trained DNNs slightly over-predict permeability values compared to no mixup.

\begin{figure}[!htbp]
  \centering
    \subfigure[No mixup:~Training prediction]
    {\includegraphics[width = 0.325\textwidth]{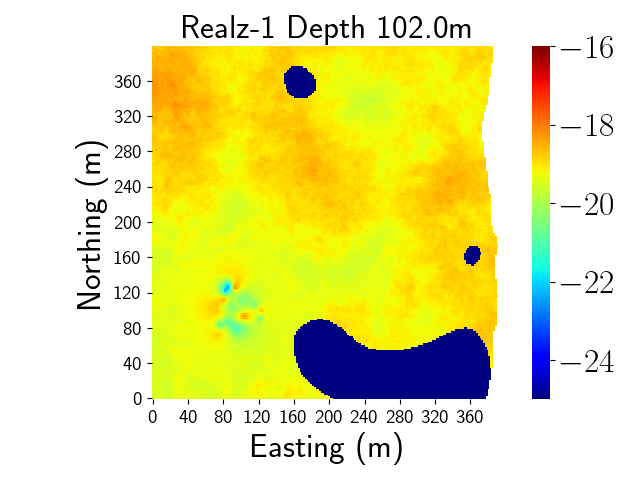}}
    \subfigure[No mixup:~Validation prediction]
    {\includegraphics[width = 0.325\textwidth]{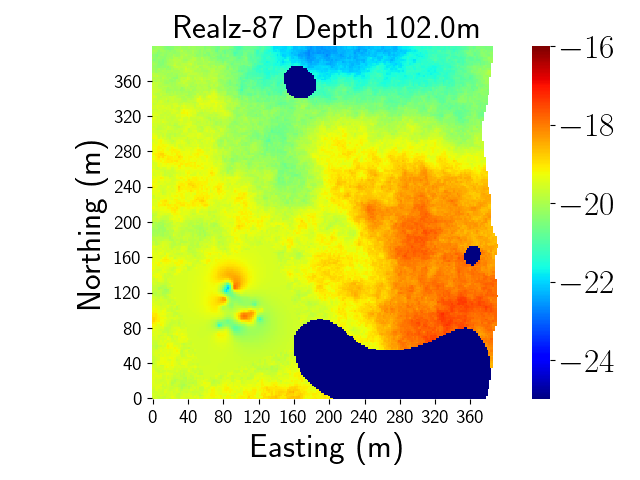}}
    \subfigure[No mixup:~Testing prediction]
    {\includegraphics[width = 0.325\textwidth]{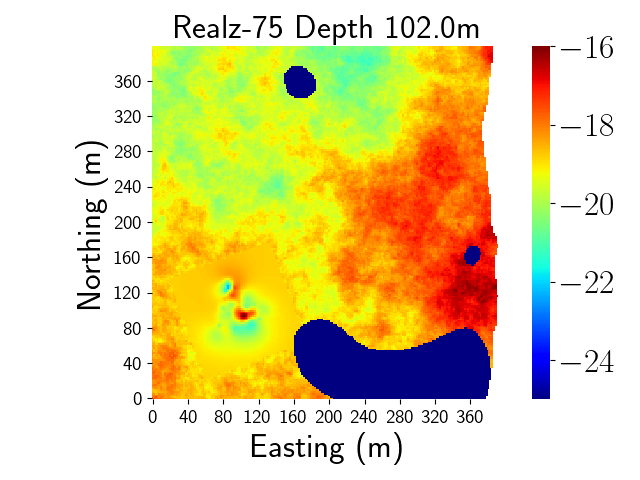}}
    \subfigure[No mixup:~Training percent errors]
    {\includegraphics[width = 0.325\textwidth]{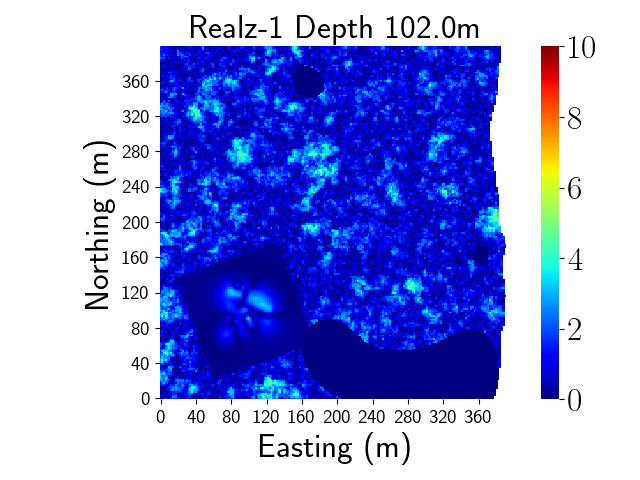}}
    \subfigure[No mixup:~Validation percent errors]
    {\includegraphics[width = 0.325\textwidth]{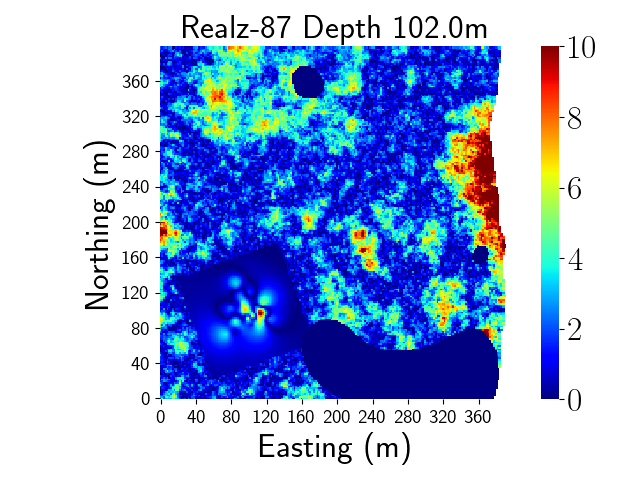}}
    \subfigure[No mixup:~Testing percent errors]
    {\includegraphics[width = 0.325\textwidth]{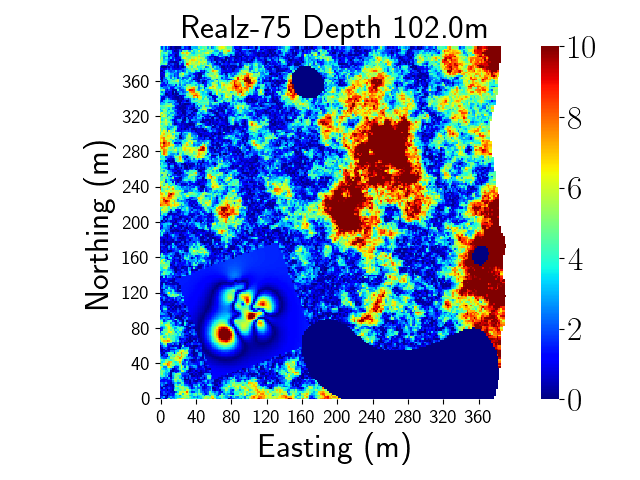}}
    \subfigure[Mixup:~Training prediction]
    {\includegraphics[width = 0.325\textwidth]{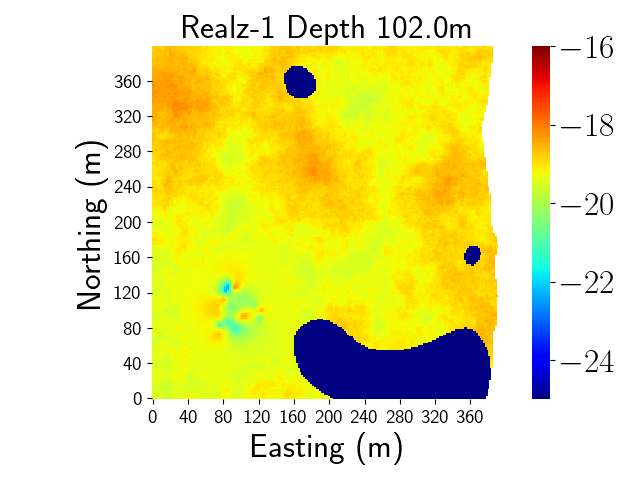}}
    \subfigure[Mixup:~Validation prediction]
    {\includegraphics[width = 0.325\textwidth]{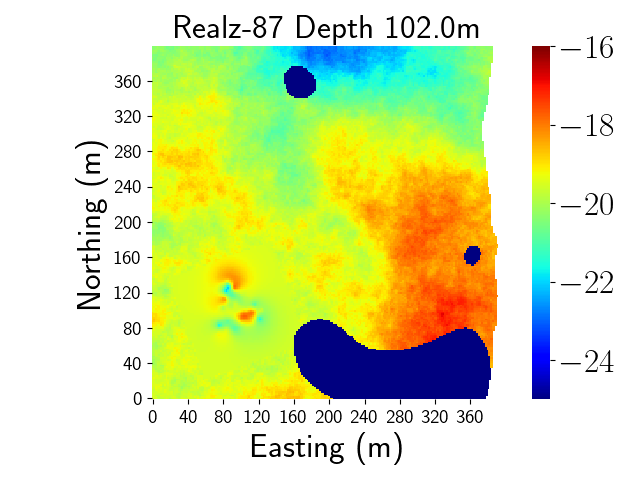}}
    \subfigure[Mixup:~Testing prediction]
    {\includegraphics[width = 0.325\textwidth]{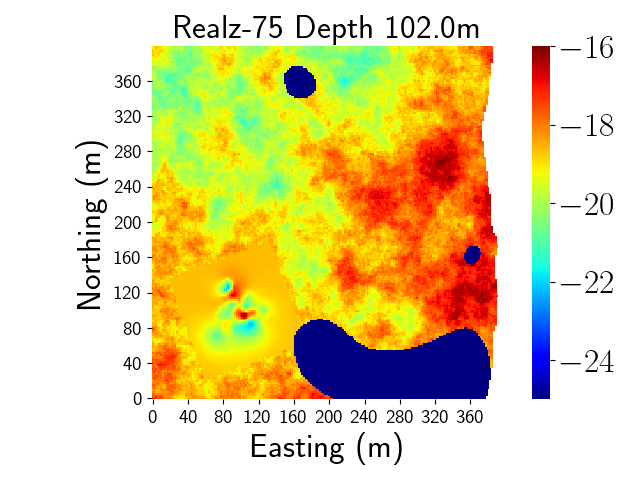}}
    \subfigure[Mixup:~Training percent errors]
    {\includegraphics[width = 0.325\textwidth]{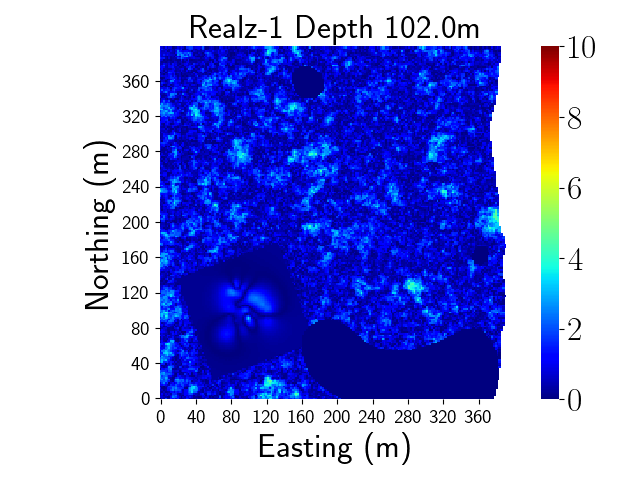}}
    \subfigure[Mixup:~Validation percent errors]
    {\includegraphics[width = 0.325\textwidth]{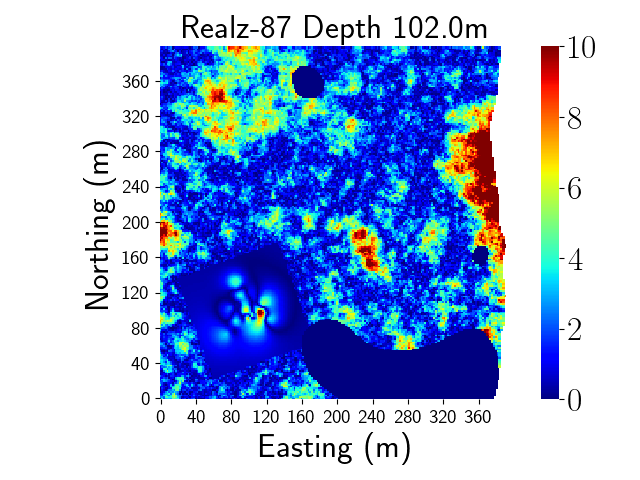}}
    \subfigure[Mixup:~Testing percent errors]
    {\includegraphics[width = 0.325\textwidth]{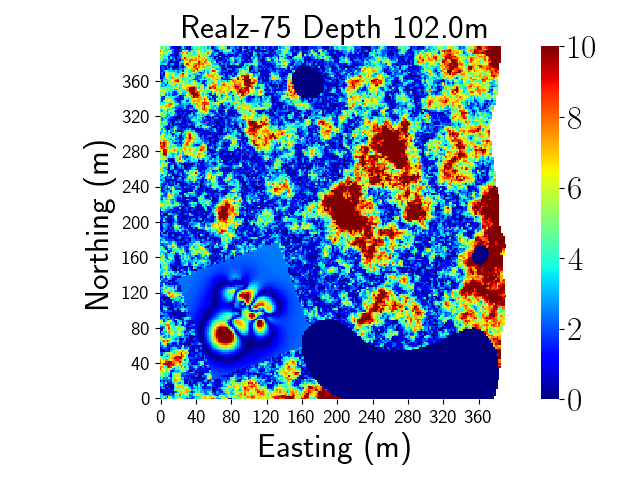}}
  \caption{\textbf{Spatial prediction and associated percent errors with and without mixup:}~This figure shows the DL-enabled inverse models (based on random seed = 0) predictions and spatial distribution of percent errors.
  The top and middle figures show the DL model predictions and associated errors without mixup, while the lower figures show pre-computed mixup after PCA. 
  The corresponding ground truth for these predictions is shown in Fig.~\ref{Fig:Perm_Realz_Data}(a), (d), and (g).
  From the percent error 2D slices, it is evident that DL models trained using mixup data perform slightly better than without mixup.
  This is because mixup-trained DL models can better predict high permeability cells than models trained without mixup.
  \label{Fig:TrainValTest_Pred_Error}}
\end{figure}

Figure~\ref{Fig:TrainValTest_Pred_Error} shows the spatial predictions and associated percent errors for tuned DNNs based on random seed = 0.
These two different DNNs are trained on data with and without mixup.
The ground truths for these predictions are shown in Fig.~\ref{Fig:Perm_Realz_Data}(a), (d), and (g).
Figures~S19-S38 in the supplementary text show the predictions of all 15 different DNNs for various training, validation, and test realizations.
Specifically, Figs.~S24-S33 show the DNN predictions for all the ten test realizations at a depth of 102m.
These supplementary figures provide predictions at fixed and varying depths ($z = 102$m to 104m) of interest.
These depths are in-between the lowest and higher groundwater levels.
We also show the spatial patterns of percent errors for specific validation and test realizations in Figs.~S39-S43.
From the main text and supplementary text figures, it is clear that models trained with and without mixup perform best on training data.
However, the predictions on validation and test realizations show that sharp features in permeability are smeared.
The spatial distribution of errors in Figs.~S39-S43 shows that certain regions along the river direction (i.e., when Easting $\approx$ 400m) have high percent errors.
But the values in most of these grid cells don't exceed 10\%.
Qualitatively, comparing spatial predictions of DNNs trained with no mixup (i.e., Figs~\ref{Fig:TrainValTest_Pred_Error}(a)-(c)) and with mixup (i.e., Figs~\ref{Fig:TrainValTest_Pred_Error}(g)-(i)) on test realizations show that mixup-enabled DNNs are able to better predict high values of permeability.
Quantitatively, the percent error plots Figs.~\ref{Fig:TrainValTest_Pred_Error}(d)-(f) and Figs.~\ref{Fig:TrainValTest_Pred_Error}(j)-(l) support this claim.
Specifically, when comparing Fig.~\ref{Fig:TrainValTest_Pred_Error}(f) and Fig.~\ref{Fig:TrainValTest_Pred_Error}(l) in regions closer to river cells (where water is flowing into subsurface), mixup has reduced percent error compared to no mixup.
This spatial pattern in errors shows that mixup-enabled models have slightly better predictive capability than models trained without mixup.
Similar inference can be made on different test realizations as shown by the figures in supplementary text (e.g., see Figs.~S24-S38 and S40-S43).

\begin{table}[htbp]
    \centering
    \caption{This table provides the overall performance of 15 different tuned DL-enabled inverse models trained on strongly and weakly labeled samples.
    }
    \resizebox{\textwidth}{!}{
    \begin{tabular}{|c|c|c|c|c|c|c|c|c|c|c|c|c|}\hline
        \multirow{2}{*}{\textbf{Metric}} &
        \multicolumn{4}{c|}{\textbf{Training (No-mixup, mixup)}} & 
        \multicolumn{4}{c|}{\textbf{Validation (No-mixup, mixup)}} & 
        \multicolumn{4}{c|}{\textbf{Testing (No-mixup, mixup)}} \\ \cline{2-13}
        & \textbf{Min.} & \textbf{Max.} & \textbf{Mean} & \textbf{SD} & \textbf{Min.} & \textbf{Max.} & \textbf{Mean} & \textbf{SD} & \textbf{Min.} & \textbf{Max.} & \textbf{Mean} & \textbf{SD} \\ \hline
        $R^2$-score & (0.18, 0.35) & (0.99, 0.99) & (0.70, 0.76) & (0.15, 0.13) & (0.14, 0.09) & (0.71, 0.71) & (0.41, 0.40) & (0.15, 0.15) & (0.03, 0.03) & (0.80, 0.81) & (0.39, 0.40) & (0.20, 0.20) \\ \hline
        MSE & (0.02, 0.01) & (0.16, 0.11) & (0.08, 0.07) & (0.02, 0.02) & (0.13, 0.12) & (0.45, 0.48) & (0.2, 0.2) & (0.08, 0.08) & (0.14, 0.14) & (1.15, 1.24) & (0.36, 0.37) & (0.27, 0.29) \\ \hline
        RMSE & (0.16, 0.11) & (0.40, 0.34) & (0.29, 0.25) & (0.03, 0.04) & (0.35, 0.35) & (0.67, 0.70) & (0.44, 0.44) & (0.08, 0.08) & (0.37, 0.37) & (1.07, 1.11) & (0.56, 0.57) & (0.20, 0.21) \\ \hline
        MAE & (0.12, 0.09) & (0.31, 0.26) & (0.22, 0.20) & (0.03, 0.03) & (0.27, 0.27) & (0.49, 0.51) & (0.34, 0.34) & (0.06, 0.06) & (0.28, 0.28) & (0.77, 0.87) & (0.42, 0.43) & (0.14, 0.15) \\ \hline
        EVS & (0.19, 0.35) & (0.99, 1.00) & (0.70, 0.76) & (0.15, 0.13) & (0.16, 0.09) & (0.71, 0.71) & (0.41, 0.40) & (0.15, 0.15) & (0.01, 0.01) & (0.8, 0.81) & (0.39, 0.40) & (0.21, 0.20) \\ \hline
        ME & (0.75, 0.52) & (3.30, 3.02) & (1.68, 1.51) & (0.36, 0.36) & (1.89, 1.87) & (3.72, 3.69) & (2.60, 2.64) & (0.49, 0.49) & (1.88, 1.90) & (4.81, 6.04) & (2.63, 2.72) & (0.84, 0.94) \\ \hline
        MAPE & (0.01, 0.00) & (0.02, 0.01) & (0.01, 0.01) & (0.001, 0.001) & (0.01, 0.01) & (0.03, 0.03) & (0.02, 0.02) & (0.001, 0.001) & (0.01, 0.01) & (0.04, 0.05) & (0.02, 0.02) & (0.01, 0.01) \\ \hline
        MedAE & (0.09, 0.07) & (0.25, 0.21) & (0.18, 0.16) & (0.02, 0.02) & (0.22, 0.22) & (0.36, 0.38) & (0.26, 0.27) & (0.04, 0.07) & (0.22, 0.22) & (0.62, 0.69) & (0.33, 0.34) & (0.10, 0.11) \\ \hline
        MPE & (4.62, 3.19) & (20.35, 18.62) & (9.10, 8.14) & (2.00, 1.93) & (9.84, 9.64) & (22.99, 23.20) & (14.52, 14.78) & (3.99, 3.90) & (9.57, 9.55) & (25.53, 30.19) & (14.29, 14.73) & (4.74, 5.03) \\ \hline
        SSIM & (0.07, 0.14) & (0.94, 0.98) & (0.49, 0.57) & (0.18, 0.18) & (0.23, 0.27) & (0.74, 0.71) & (0.46, 0.46) & (0.12, 0.11) & (0.26, 0.24) & (0.66, 0.66) & (0.33, 0.43) & (0.10, 0.08) \\ \hline
    \end{tabular}}
    \label{table:Performance_Metrics_10}
\end{table}

The performance metrics of tuned DNNs with and without mixup are shown in Table~\ref{table:Performance_Metrics_10}.
The minimum, maximum, mean, and standard deviation (SD) are evaluated on the ensemble predictions -- $282$ training realizations $\times 15$ model predictions, $10$ validation realizations $\times 15$ model predictions, and $10$ testing realizations $\times 15$ model predictions.
Figures~S44-S48 provide detailed plots of each metric for 282 training (strongly labeled), ten validation, and ten testing realizations using 15 different tuned models. 
These metrics \cite{scikit-learn} are computed on the entire 3D $\ln[k(\mathbf{x})]$ field.
Each metrics captures specific local or global patterns in the predictions.
For example, $R^2$-score illustrates the goodness of fit, which measures how well the 3D permeability field is likely to be predicted by the tuned DL-enabled inverse models.
Mean squared error (MSE), and root mean squared error (RMSE) are scale-dependent accuracy metrics, which are sensitive to outliers.
MSE and RMSE measure the squares of errors and the square root of the average of squared errors, respectively.
Mean absolute error (MAE) provides the average of the absolute values of the errors.
Explained variance score (EVS) measures the magnitude by which a tuned DL-enabled inverse model accounts for the variation of a given realization.
Maximum error (ME) provides the worst-case error between the predicted and the true $\ln[k(\mathbf{x})]$ values.
Mean absolute percentage error (MAPE) calculates the relative percentage error with respect to actual value of $\ln[k(\mathbf{x})]$.
MAPE is not scale-dependent and provides information on the sensitivity of DL-enabled inverse model predictions to relative errors.
Median absolute error (MedAE) is robust to outliers and is calculated by taking the median of all absolute differences between the ground truth and predictions.
It tells us about the variability in $\ln[k(\mathbf{x})]$ predictions by DL-enabled inverse models.
Maximum percent error (MPE) estimates the maximum possible percent error in predictions.
Structural similarity index measure (SSIM) incorporates structural information and indicates how well the DL-enabled inverse models capture spatial patterns at various depths.
It measures the similarity between 2D slices of ground truth and predictions.
The overall metric values (e.g., mean, standard deviation in Table~\ref{table:Performance_Metrics_10}) on training and validation sets show that DNNs trained with and without mixup have similar performance.
But on the testing set, we can see a slight improvement in the performance of DNN tuned using mixup data.
For example, the SSIM scores using mixup-enabled DNNs are slightly better than tuned models trained without mixup.

\begin{figure}
  \centering
    \subfigure[$R^2$-score]
    {\includegraphics[width = 0.31\textwidth]{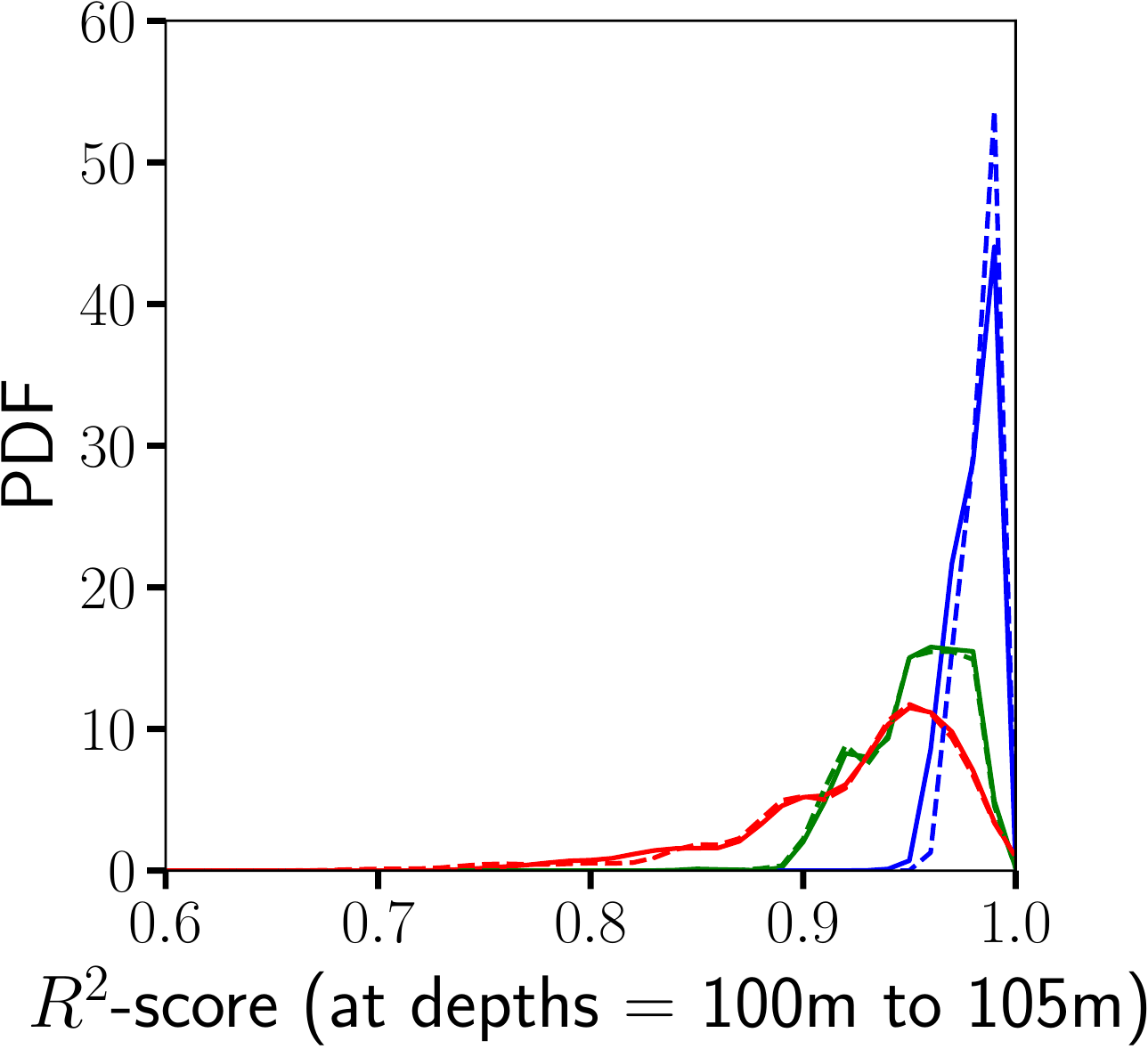}}
    \hspace{0.1in}
    \subfigure[SSIM]
    {\includegraphics[width = 0.29\textwidth]{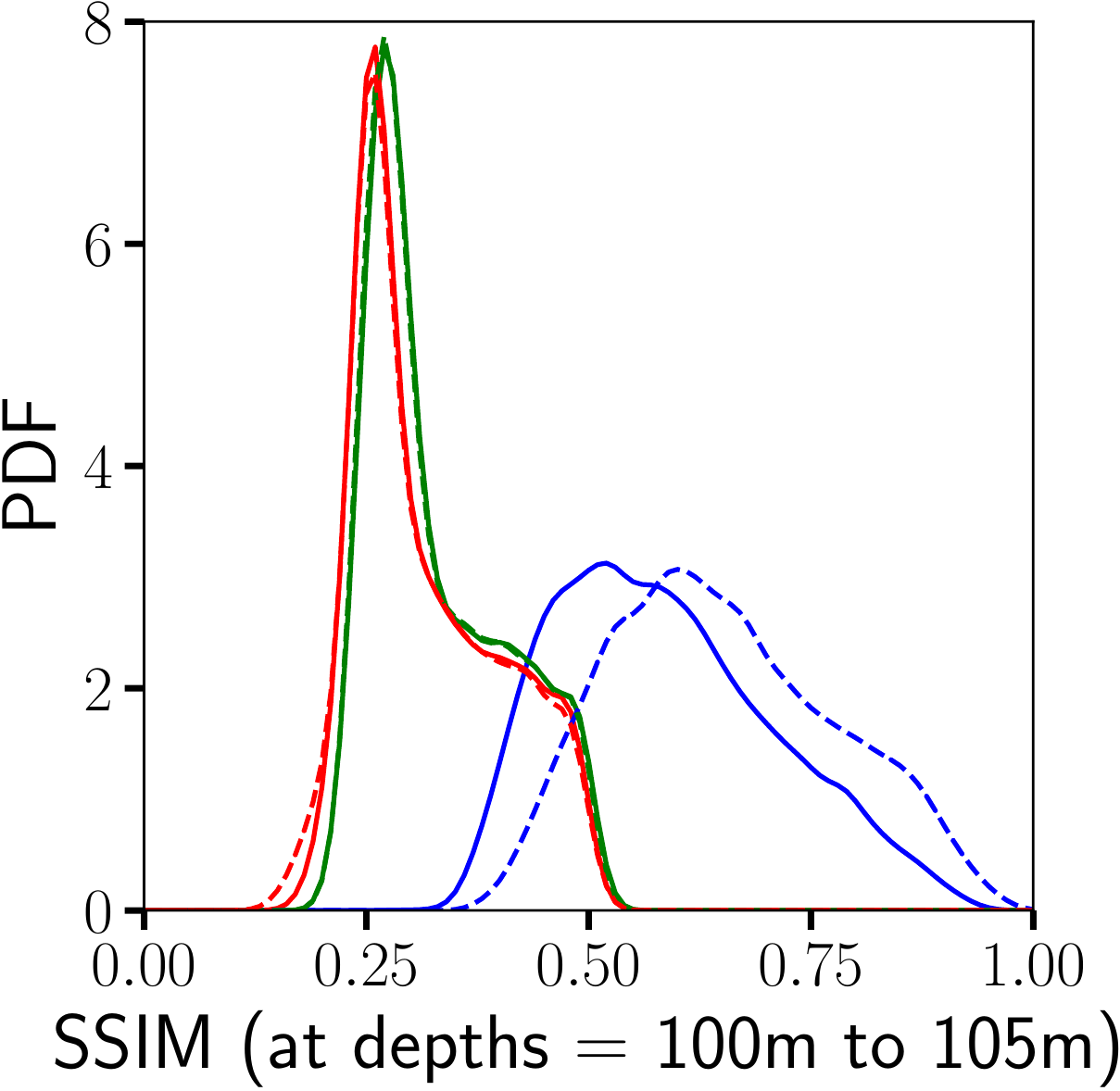}}
    \hspace{0.1in}
    \subfigure[Percent error in each grid cell]
    {\includegraphics[width = 0.28\textwidth]{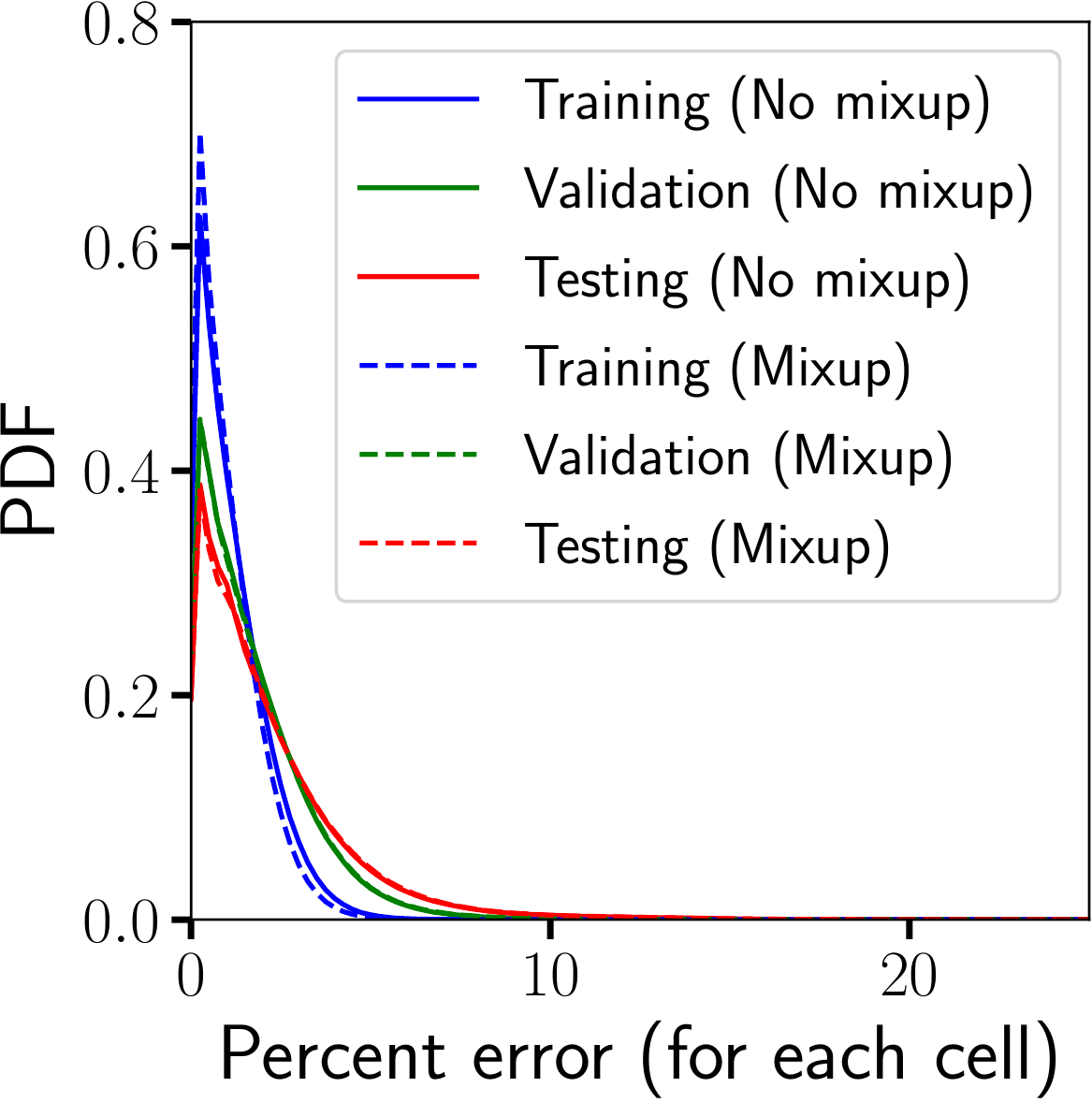}}
  \caption{\textbf{Distribution of $R^2$-score, SSIM metric, and percent errors for with and without mixup:}~These figures compares the probability density estimates of all the 15 different tuned model trained with and without mixup.
  Metrics based on DL model predictions are evaluated at depths between 100m to 105m, the zone of interest.
  This zone is between the lowest and highest groundwater levels.
  The left and middle figures show the $R^2$-score and SSIM metric values, while the right figure shows the percent error in each cell.
  From $R^2$-score, quantitatively, it is evident that DL model predictions can capture the overall trends in the 3D permeability field.
  Qualitatively, it is clear that the percent errors in each cell are less than 10\%, and density function is concentrated primarily with 3\%.
  However, the SSIM score reveals reduced performance in capturing local patterns, which shows that over-fitting is still prevalent.
  In all these scenarios, the mixup trained model performs slightly better than no mixup.
  \label{Fig:TrainValTest_R2_SSIM}}
\end{figure}

Figure~\ref{Fig:TrainValTest_R2_SSIM}(a) and (b) shows the PDF values of $R^2$-score and SSIM at depths of interest.
These depths are between 100m to 105m, where river water intrudes and mixes with existing groundwater.
To compute the $R^2$-score and SSIM at these depths, we include the Ringold and river cell permeability values in our metric estimations.
As a result, we have an artificial increase in $R^2$-score and SSIM values at these depths. 
For example, we can see $R^2$-score's probability density is concentrated between 0.75 and 1.0.
We can draw similar inferences on SSIM metric values.
However, from this SSIM figure, it is evident that mixup-enabled DNNs can better capture the spatial patterns in the permeability field than models trained without mixup.
Figure~\ref{Fig:TrainValTest_R2_SSIM}(c) shows the PDFs of percent errors in each grid cell for all depths (i.e., from 90m to 110m).
The PDF vs. percent errors also provides essential insight that DNN can estimate spatially distributed permeability within 10\% error even under limited simulation data as proability density is primarily concentrated within this error.

\subsection{\textbf{Discussion}}
\label{SubSec:S4_Discussion}
Our results demonstrate the applicability of using deep learning to estimate 3D permeability using time-lapse ERT.
However, there are certain limitations of the current methodology, primarily related to reconstruction errors and over-fitting under limited simulation data.
Herein, we briefly describe possible approaches to overcome these limitations to capture spatial patterns better.
Our next step is to test our enhanced DL methodology to observational errors and estimate 3D permeability using real time-lapse ERT field data.
\begin{enumerate}
  \item \textit{Data pre-processing}:~Typically, DL algorithms perform better when standardized using advanced pre-processing methods \cite{garcia2015data} such as robust scalers or custom transformers (e.g., power transform).
  A detailed comparison of various transformations to pre-process ERT simulations for dimensionality reduction might shed better insights on the predictive capability of trained DNNs.
  \item \textit{Generating targeted realizations}:~Due to the curse of dimensionality, the number of training samples needed to reduce the over-fitting of DNNs can be huge.
  However, as discussed in Sec.~\ref{SubSec:S4_Comp_Cost}, generating training data is expensive and requires HPC resources.
  To obtain the best DNN performance, generating targeted training samples is necessary.
  Recent advances in active learning \cite{smith2018less} can provide us with better algorithms to explore and sample the high-dimensional permeability parameter space efficiently.
  \item \textit{Hydrogeophysics data streams}:~Our proposed method ingests time-lapse ERT data to estimate permeability.
  To better constrain this 3D field, we can extend our existing workflow to incorporate flow, transport, electrical conductivity, and temperature data as inputs.
  In such a scenario, a tuned DNN takes multiple PCA components, where each input layer corresponds to flow, transport, and time-lapse ERT encoded variables.
  \item \textit{Advanced mixup strategies}:~The current study used classical Mixup data augmentation on latent spaces, which produced smaller performance gain over no mixup.
  We note that classical mixup uses linearity assumption, which does not necessarily hold in a regression setting.
  This is because the label space is continuous, and taking a linear interpolation of training examples that are very different (e.g., data-wise, label-wise) may produce incorrect labels.
  Advanced mixup strategies (e.g., Puzzle Mix, multi-dimensional mixup, MixRL) \cite{kim2020puzzle,hwang2021mixrl} can be employed to temper DNN confidence and achieve a significant reduction in model over-fitting.
  For example, the MixRL method accounts for the underlying statistics of the training data and can learn how many nearest neighbors we should mix a sample with using reinforcement learning.
  This importance of data valuation using reinforcement learning limits the number of nearest neighbors to mix a given sample instead of mixing with all combinations. 
  \item \textit{Few-shot learning}:~Developing traditional DNNs requires substantial training data.
  Recent advances in few-shot learning \cite{wang2019few} can be used to build accurate models.
  These approaches use prior knowledge about similarity and learn patterns in training data, thereby constraining the learning algorithm to choose parameters that generalize well from few examples.
  \item \textit{Improving permeability resolution}:~From the DNN predictions, it is clear that estimated permeability is smeared.
  We can improve the estimated permeability resolution by correcting it through an autoencoder trained using Mixup
  \item \textit{Model averaging and stacking to improve predictions}:~In the current DL methodology, we developed 30 different tuned models based on popular random seeds.
  Each trained model is then used for permeability estimation.
  We can improve the predictive performance of these models by combining the predictions from all the trained DNNs.
  We can achieve this through ensemble deep learning \cite{ganaie2021ensemble}, which not only reduces the variance of permeability predictions but also can result in estimations that are better than any single DL model.
  \item \textit{Advanced dimensionality reduction to account for 3D patterns}:~In our study, we flatten the 3D tensor to a 1D vector to perform dimensionality reduction.
  Such a transformation may not always capture patterns across depth of the permeability field.
  We can employ tensor factorization methods \cite{cichocki2009nonnegative} to overcome such a limitation and better capture spatial patterns.
  For example, non-negative tensor factorization \cite{vesselinov2019unsupervised} allows us to directly work with the 3D tensor to extract latent variables representing spatial features across depth.
  This enhanced dimensionality reduction may allow for development of better DNNs to map encoded ERT to encoded permeability.
  \item \textit{Permeability estimation with predictive uncertainty}:~Our proposed DNNs are deterministic and do not provide uncertainty in their prediction.
  We can improve our workflow to accommodate Bayesian neural networks to map encoded ERT to the encoded permeability field for developing probabilistic ML models.
  \item \textit{Embedding knowledge into neural network architectures}:~Incorporating prior information and domain knowledge through Knowledge-guided machine learning \cite{raissi2019physics,li2020fourier} may allow to reduce over-fitting.
  For example, we can train neural networks by embedding the governing laws to simulate hydrogeophysics for our problem.
  This embedding of physics allows us to solve forward and inverse problems in one shot.
  However, various practical challenges may need to be addressed to develop such a workflow \cite{karra2021adjointnet}.
\end{enumerate}

\section{\textbf{CONCLUSIONS}}
\label{Sec:S5_Conclusions}
We have developed a fast and reasonably accurate DL methodology to invert permeability under limited training samples.
{\color{black}Our DL approach combines dimensionality reduction methods with deep neural networks to map time-lapse characteristics of encoded ERT measurements to encoded 3D permeability field.}
PCA-based encoder-decoders are trained to perform dimensionality reduction.
The extracted principal components from potential and permeability fields are used to train DNNs.
We have performed hyperparameter tuning using the Bayesian optimization method to identify optimal DNN architectures.
The data used for training the DNNs and PCA are based on hydrogeophysics simulations using process models in \texttt{PFLOTRAN} and \texttt{E4D}.
However, training DNNs under limited simulation data is challenging as it resulted in model over-fitting.
Mixup-based data augmentation was used to generate additional training samples to improve DNN performance.
Results (e.g., performance metric based on SSIM) showed that mixup-based DNNs perform slightly better than no-mixup, but over-fitting still exists.
{\color{black}Probability density estimates provided insights on high permeable regions (between z = 100m to 105m), where the percent errors in each grid cell are high ($\approx10\%$).}
Overall, the DNNs trained on the proposed methodology can capture spatial patterns in permeability within an error of 10\%.
{\color{black}Finally, we note that the computational cost to train a tuned encoded DNNs is minimal ($\approx$ 1 processor hour on a laptop CPU) compared to running a single forward simulation ($\approx$ 2500 processor hours on a NERSC's CPU).}
This fast inference with justifiable predictive capability makes our DL methodology reasonably attractive for solving high-dimensional inverse problems under limited training data.
{\color{black}We acknowledge that there are certain limits to this fast inference. DL inference and predictions are reasonably accurate for low permeability regions. However, we have to be cautious in inferring areas of high permeability (e.g., closer to the river) due to the reduced performance (i.e., over-fitting) of DL models.}
Our future work involves improving the predictive capability of our DL workflow.
This improvement can be achieved in multiple ways through few-shot learning, Bayesian neural networks to account for uncertainty, and active learning to generate targeted training data.

\section*{\textbf{ACKNOWLEDGMENTS}}
This work was funded by the ExaSheds project, which was supported by the United States Department of Energy, Office of Science, Office of Biological and Environmental Research, Earth and Environmental Systems Sciences Division, Data Management Program, under Award Number DE-AC02-05CH11231.
This research study also used the National Energy Research Scientific Computing Center (NERSC) resources, a U.S. Department of Energy Office of Science User Facility located at Lawrence Berkeley National Laboratory, operated under Contract No. DE-AC02-05CH11231.
The authors' views and opinions expressed herein do not necessarily state or reflect those of the United States Government or any agency thereof.

\section*{ABBREVIATIONS}
\begin{itemize}
  \item DNN:~Deep Neural Network
  \item DL:~Deep Learning
  \item \texttt{E4D}:~A scalable, 3D geophysical modeling and inversion code
  \item ERT:~Electrical Resistivity Tomography
  \item EVR:~Explained Variance Ratio
  \item EVS:~Explained Variance Score
  \item FAIR:~Findable Accessible Interoperable and Reusable
  \item GPR:~Gaussian Process Regression
  \item GP-UCB:~Gaussian Process Upper Confidence Bound
  \item HPC:~High-Performance Computing
  \item KDE: Kernel Density Estimation
  \item ML:~Machine Learning
  \item MAE:~Mean Absolute Error
  \item MAPE:~Mean Absolute Percent Error
  \item MedAE:~Median Absolute Error
  \item ME:~Maximum Error
  \item MPE:~Maximum Percent Error
  \item MSE:~Mean Squared Error
  \item NERSC:~National Energy Research Scientific Computing Center
  \item PCA:~Principal Component Analysis
  \item PDF:~Probability Density Function
  \item PE:~Percent Error
  \item \texttt{PFLOTRAN}:~An open source, state-of-the-art massively parallel subsurface flow and reactive transport code
  \item RMSE:~Root Mean Squared Error
  \item SD:~Standard Deviation
  \item SSIM:~Structural Similarity Index Measure
\end{itemize}

\section*{\textbf{CONFLICT OF INTEREST}}
The authors declare that they do not have any conflicts of interest.

\section*{\textbf{APPENDIX (supplementary materials)}}
Supplementary materials and associated figures can be found in a separate file.

\section*{\textbf{DATA, CODE AVAILABILITY, AND RESOURCES}}
The data generated for the proposed deep learning model development uses open-science principles.
Specifically, we make the data Findable Accessible Interoperable and Reusable (FAIR).
FAIR principles expedite community-based data generation, modeling, interdisciplinary collaboration and provides a means to test new hypotheses.
The datasets generated and analyzed, as well as the scripts for this study will be made available on this Github repository:~\url{https://github.com/maruti-iitm/DL4HGP.git}.
\texttt{PFLOTRAN} source code can be downloaded at \url{https://bitbucket.org/pflotran/pflotran/src/master/}.
\texttt{E4D} source code can be downloaded at \url{https://github.com/pnnl/E4D}.
\bibliographystyle{unsrt}
\typeout{} 
\bibliography{Master_References/HGP_DL_References}

\begin{thebibliography}{100}

\bibitem{al2004hydro}
S.~A. Hagrey, R.~Meissner, U.~Werban, W.~Rabbel, and A.~Ismaeil.
\newblock Hydro-, bio-geophysics.
\newblock {\em The Leading Edge}, 23:670--674, 2004.

\bibitem{atekwana2009biogeophysics}
E.~A. Atekwana and L.~D. Slater.
\newblock {Biogeophysics:~A new frontier in Earth science research}.
\newblock {\em Reviews of Geophysics}, 47, 2009.

\bibitem{chen2020integrating}
X.~Chen, R.~M. Lee, D.~Dwivedi, K.~Son, Y.~Fang, X.~Zhang, E.~Graham,
  J.~Stegen, J.~B. Fisher, D.~Moulton, and T.~D. Scheibe.
\newblock Integrating field observations and process-based modeling to predict
  watershed water quality under environmental perturbations.
\newblock {\em Journal of Hydrology}, page 125762, 2020.

\bibitem{mudunuru2021subsurface}
M.~K. Mudunuru, J.~W. Carey, L.~Chen, Q.~Kang, S.~Karra, V.~V. Vesselinov,
  R.~S. Middleton, P.~A. Johnson, and H.~S. Viswanathan.
\newblock {Subsurface Energy:~Flow and Reactive-Transport in Porous and
  Fractured Media}.
\newblock In {\em Handbook of Porous Materials: Synthesis, Properties, Modeling
  and Key Applications Volume 4-Porous Materials for Energy Conversion and
  Storage}, pages 323--395. World Scientific, 2021.

\bibitem{berkowitz2002characterizing}
B.~Berkowitz.
\newblock {Characterizing flow and transport in fractured geological media:~A
  review}.
\newblock {\em Advances in Water Resources}, 25:861--884, 2002.

\bibitem{neuman2005trends}
S.~P. Neuman.
\newblock Trends, prospects and challenges in quantifying flow and transport
  through fractured rocks.
\newblock {\em Hydrogeology Journal}, 13:124--147, 2005.

\bibitem{allen2007microbial}
J.~P. Allen, E.~A. Atekwana, E.~A. Atekwana, J.~W. Duris, D.~D. Werkema, and
  S.~Rossbach.
\newblock The microbial community structure in petroleum-contaminated sediments
  corresponds to geophysical signatures.
\newblock {\em Applied and Environmental Microbiology}, 73:2860--2870, 2007.

\bibitem{atekwana2010geophysical}
E.~A. Atekwana and E.~A. Atekwana.
\newblock {Geophysical signatures of microbial activity at hydrocarbon
  contaminated sites:~A review}.
\newblock {\em Surveys in Geophysics}, 31:247--283, 2010.

\bibitem{middleton2015shale}
R.~S. Middleton, J.~W. Carey, R.~P. Currier, J.~D. Hyman, Q.~Kang, S.~Karra,
  J.~J.-Mart{\'\i}nez, M.~L. Porter, and H.~S. Viswanathan.
\newblock {Shale gas and non-aqueous fracturing fluids:~Opportunities and
  challenges for supercritical $\mathrm{C0}_2$}.
\newblock {\em Applied Energy}, 147:500--509, 2015.

\bibitem{hyman2016understanding}
J.~D. Hyman, J.~J.-Mart{\'\i}nez, H.~S. Viswanathan, J.~W. Carey, M.~L. Porter,
  E.~Rougier, S.~Karra, Q.~Kang, L.~Frash, and L.~Chen.
\newblock {Understanding hydraulic fracturing:~A multi-scale problem}.
\newblock {\em Philosophical Transactions of the Royal Society A: Mathematical,
  Physical and Engineering Sciences}, 374:20150426, 2016.

\bibitem{mudunuru2020physics}
M.~K. Mudunuru, D.~O'Malley, S.~Srinivasan, J.~D. Hyman, M.~Ryan Sweeney, L.~P.
  Frash, J.~W. Carey, M.~Robert Gross, N.~J. Welch, and S.~Karra.
\newblock {Physics-informed machine learning for real-time unconventional
  reservoir management}.
\newblock {\em AAAI-MLPS Conference}, 2020.

\bibitem{stauffer2009system}
P.~H. Stauffer, H.~S. Viswanathan, R.~J. Pawar, and G.~D. Guthrie.
\newblock A system model for geologic sequestration of carbon dioxide, 2009.

\bibitem{middleton2012cross}
R.~S. Middleton, G.~N. Keating, P.~H. Stauffer, A.~B. Jordan, H.~S.
  Viswanathan, Q.~J. Kang, J.~W. Carey, M.~L. Mulkey, E.~J. Sullivan, and S.~P.
  Chu.
\newblock {The cross-scale science of $\mathrm{CO}_2$ capture and storage:~From
  pore scale to regional scale}.
\newblock {\em Energy \& Environmental Science}, 5:7328--7345, 2012.

\bibitem{ahmmed2021machine}
B.~Ahmmed, S.~Karra, V.~V. Vesselinov, and M.~K. Mudunuru.
\newblock {Machine learning to discover mineral trapping signatures due to
  $\mathrm{CO}_2$ injection}.
\newblock {\em International Journal of Greenhouse Gas Control}, 109:103382,
  2021.

\bibitem{rutqvist2009comparative}
J.~Rutqvist, D.~Barr, J.~T. Birkholzer, K.~Fujisaki, O.~Kolditz, Q.-S. Liu,
  T.~Fujita, W.~Wang, and C.-Y. Zhang.
\newblock {A comparative simulation study of coupled THM processes and their
  effect on fractured rock permeability around nuclear waste repositories}.
\newblock {\em Environmental Geology}, 57:1347--1360, 2009.

\bibitem{apted2017geological}
M.~J. Apted and J.~Ahn.
\newblock {\em {Geological Repository Systems for Safe Disposal of Spent
  Nuclear Fuels and Radioactive Waste}}.
\newblock Woodhead Publishing, 2017.

\bibitem{brown2012mining}
D.~W. Brown, D.~V. Duchane, G.~Heiken, and V.~T. Hriscu.
\newblock {\em {Mining the Earth's heat:~Hot Dry Rock Geothermal Energy}}.
\newblock Springer Science \& Business Media, 2012.

\bibitem{mudunuru2017regression}
M.~K. Mudunuru, S.~Karra, D.~R. Harp, G.~D. Guthrie, and H.~S. Viswanathan.
\newblock Regression-based reduced-order models to predict transient thermal
  output for enhanced geothermal systems.
\newblock {\em Geothermics}, 70:192--205, 2017.

\bibitem{siler2021machine}
D.~L. Siler, J.~D. Pepin, V.~V. Vesselinov, M.~K. Mudunuru, and B.~Ahmmed.
\newblock {Machine learning to identify geologic factors associated with
  production in geothermal fields:~A case-study using 3D geologic data, Brady
  geothermal field, Nevada}.
\newblock {\em Geothermal Energy}, 9:1--17, 2021.

\bibitem{SMART_CS_Initiative}
{SMART Initiative:~Science-informed Machine Learning for Accelerating Real-Time
  Decisions in Subsurface Applications}.
\newblock \url{{https://edx.netl.doe.gov/smart/}}, 2021.

\bibitem{guerin2005borehole}
R.~Gu{\'e}rin.
\newblock Borehole and surface-based hydrogeophysics.
\newblock {\em Hydrogeology Journal}, 13:251--254, 2005.

\bibitem{turk2011subsurface}
A.~S. Turk, K.~A. Hocaoglu, and A.~A. Vertiy.
\newblock {\em Subsurface Sensing}, volume 197.
\newblock John Wiley \& Sons, 2011.

\bibitem{saleh2011introduction}
B.~Saleh.
\newblock {\em Introduction to Subsurface Imaging}.
\newblock Cambridge University Press, 2011.

\bibitem{preko2009comparison}
K.~Preko, A.~Scheuermann, and H.~Wilhelm.
\newblock {Comparison of invasive and non-invasive electromagnetic methods in
  soil water content estimation of a Dike model}.
\newblock {\em Journal of Geophysics and Engineering}, 6:146--161, 2009.

\bibitem{binley2015emergence}
A.~Binley, S.~S. Hubbard, J.~A. Huisman, A.~Revil, D.~A. Robinson, K.~Singha,
  and L.~D. Slater.
\newblock The emergence of hydrogeophysics for improved understanding of
  subsurface processes over multiple scales.
\newblock {\em Water Resources Research}, 51:3837--3866, 2015.

\bibitem{singha2015advances}
K.~Singha, F.~D. Day-Lewis, T.~Johnson, and L.~D. Slater.
\newblock Advances in interpretation of subsurface processes with time-lapse
  electrical imaging.
\newblock {\em Hydrological Processes}, 29:1549--1576, 2015.

\bibitem{parsekian2015multiscale}
A.~D. Parsekian, K.~Singha, B.~J. Minsley, W.~S. Holbrook, and L.~Slater.
\newblock Multiscale geophysical imaging of the critical zone.
\newblock {\em Reviews of Geophysics}, 53:1--26, 2015.

\bibitem{misra2019deep}
S.~Misra and H.~Li.
\newblock Deep neural network architectures to approximate the fluid-filled
  pore size distributions of subsurface geological formations.
\newblock {\em Machine Learning for Subsurface Characterization}, 183, 2019.

\bibitem{li2020neural}
H.~Li, S.~Misra, and J.~He.
\newblock Neural network modeling of in situ fluid-filled pore size
  distributions in subsurface shale reservoirs under data constraints.
\newblock {\em Neural Computing and Applications}, 32:3873--3885, 2020.

\bibitem{slater2007near}
L.~Slater.
\newblock {Near surface electrical characterization of hydraulic
  conductivity:~From petrophysical properties to aquifer geometries--A review}.
\newblock {\em Surveys in Geophysics}, 28:169--197, 2007.

\bibitem{robinson2008advancing}
D.~A. Robinson, A.~Binley, N.~Crook, F.~D. Day-Lewis, T.~P.~A. Ferr{\'e},
  V.~J.~S. Grauch, R.~Knight, M.~Knoll, V.~Lakshmi, and R.~Miller.
\newblock {Advancing process-based watershed hydrological research using
  near-surface geophysics:~A vision for, and review of, electrical and magnetic
  geophysical methods}.
\newblock {\em Hydrological Processes: An International Journal},
  22:3604--3635, 2008.

\bibitem{rubin2006hydrogeophysics}
Y.~Rubin and S.~S. Hubbard.
\newblock {\em Hydrogeophysics}, volume~50.
\newblock Springer Science \& Business Media, 2006.

\bibitem{seo2012nonlinear}
J.~K. Seo and E.~J. Woo.
\newblock {\em Nonlinear Inverse Problems in Imaging}.
\newblock John Wiley \& Sons, 2012.

\bibitem{mueller2012linear}
J.~L. Mueller and S.~Siltanen.
\newblock {\em Linear and Nonlinear Inverse Problems with Practical
  Applications}.
\newblock SIAM, 2012.

\bibitem{sen2013global}
M.~K. Sen and P.~L. Stoffa.
\newblock {\em Global Optimization Methods in Geophysical Inversion}.
\newblock Cambridge University Press, 2013.

\bibitem{tarantola2005inverse}
A.~Tarantola.
\newblock {\em Inverse Problem Theory and Methods for Model Parameter
  Estimation}.
\newblock SIAM, 2005.

\bibitem{aster2018parameter}
R.~C. Aster, B.~Borchers, and C.~H. Thurber.
\newblock {\em Parameter Estimation and Inverse Problems}.
\newblock Elsevier, 2018.

\bibitem{chen2013application}
X.~Chen, G.~E. Hammond, C.~J. Murray, M.~L. Rockhold, V.~R. Vermeul, and J.~M.
  Zachara.
\newblock {Application of ensemble-based data assimilation techniques for
  aquifer characterization using tracer data at Hanford 300 area}.
\newblock {\em Water Resources Research}, 49:7064--7076, 2013.

\bibitem{asch2016data}
M.~Asch, M.~Bocquet, and M.~Nodet.
\newblock {\em Data Assimilation:~Methods, Algorithms, and Applications}.
\newblock SIAM, 2016.

\bibitem{jiang2021dart}
P.~Jiang, X.~Chen, K.~Chen, J.~Anderson, N.~Collins, and M.~E. Gharamti.
\newblock {DART-PFLOTRAN:~An ensemble-based data assimilation system for
  estimating subsurface flow and transport model parameters}.
\newblock {\em Environmental Modelling \& Software}, 142:105074, 2021.

\bibitem{doherty2010approaches}
J.~E. Doherty and R.~J. Hunt.
\newblock {\em {Approaches to highly parameterized inversion:~A guide to using
  PEST for groundwater-model calibration}}, volume 2010.
\newblock US Department of the Interior, US Geological Survey Reston, 2010.

\bibitem{anderson2009data}
J.~Anderson, T.~Hoar, K.~Raeder, H.~Liu, N.~Collins, R.~Torn, and A.~Avellano.
\newblock {The data assimilation research testbed:~A community facility}.
\newblock {\em Bulletin of the American Meteorological Society}, 90:1283--1296,
  2009.

\bibitem{adams2009dakota}
B.~M. Adams, W.~J. Bohnhoff, K.~R. Dalbey, J.~P. Eddy, M.~S. Eldred, D.~M. Gay,
  K.~Haskell, P.~D. Hough, and L.~P. Swiler.
\newblock {DAKOTA, a multilevel parallel object-oriented framework for design
  optimization, parameter estimation, uncertainty quantification, and
  sensitivity analysis:~Version 5.0 user's manual}.
\newblock {\em Sandia National Laboratories, Tech. Rep. SAND2010-2183}, 2009.

\bibitem{stuart2010inverse}
A.~M. Stuart.
\newblock {Inverse problems:~A Bayesian perspective}.
\newblock {\em Acta numerica}, 19:451--559, 2010.

\bibitem{mudunuru2017sequential}
M.~K. Mudunuru, S.~Karra, N.~Makedonska, and T.~Chen.
\newblock Sequential geophysical and flow inversion to characterize fracture
  networks in subsurface systems.
\newblock {\em Statistical Analysis and Data Mining: The ASA Data Science
  Journal}, 10:326--342, 2017.

\bibitem{maclaren2019can}
O.~J. Maclaren and R.~Nicholson.
\newblock {What can be estimated? Identifiability, estimability, causal
  inference and ill-posed inverse problems}.
\newblock {\em arXiv preprint arXiv:1904.02826}, 2019.

\bibitem{caers2011modeling}
J.~Caers.
\newblock {\em Modeling Uncertainty in the Earth Sciences}.
\newblock John Wiley \& Sons, 2011.

\bibitem{scheidt2018quantifying}
C.~Scheidt, L.~Li, and J.~Caers.
\newblock {\em Quantifying Uncertainty in Subsurface systems}, volume 236.
\newblock John Wiley \& Sons, 2018.

\bibitem{hu2008multiple}
L.~Y. Hu and T.~Chugunova.
\newblock Multiple-point geostatistics for modeling subsurface heterogeneity:~a
  comprehensive review.
\newblock {\em Water Resources Research}, 44, 2008.

\bibitem{oware2013physically}
E.~K. Oware, S.~M.~J. Moysey, and T.~Khan.
\newblock Physically based regularization of hydrogeophysical inverse problems
  for improved imaging of process-driven systems.
\newblock {\em Water Resources Research}, 49:6238--6247, 2013.

\bibitem{sun2019can}
A.~Y. Sun and B.~R. Scanlon.
\newblock {How can Big Data and machine learning benefit environment and water
  management:~A survey of methods, applications, and future directions}.
\newblock {\em Environmental Research Letters}, 14:073001, 2019.

\bibitem{tahmasebi2020machine}
P.~Tahmasebi, S.~Kamrava, T.~Bai, and M.~Sahimi.
\newblock {Machine learning in geo-and environmental sciences:~From small to
  large scale}.
\newblock {\em Advances in Water Resources}, 142:103619, 2020.

\bibitem{jagtap2021deep}
N.~V. Jagtap, M.~K. Mudunuru, and K.~B. Nakshatrala.
\newblock A deep learning modeling framework to capture mixing patterns in
  reactive-transport systems.
\newblock {\em Communications in Computational Physics (arXiv preprint
  arXiv:2101.04227) DOI:~0.4208/cicp.OA-2021-0088}, 2021.

\bibitem{xu2021machine}
T.~Xu and F.~Liang.
\newblock {Machine learning for hydrologic sciences:~An introductory overview}.
\newblock {\em Wiley Interdisciplinary Reviews: Water}, page e1533, 2021.

\bibitem{cromwell2021estimating}
E.~L.~D. Cromwell, P.~Shuai, P.~Jiang, E.~Coon, S.~L. Painter, D.~Moulton,
  Y.~Lin, and X.~Chen.
\newblock Estimating watershed subsurface permeability from stream discharge
  data using deep neural networks.
\newblock {\em Frontiers in Earth Science}, 9, 2021.

\bibitem{sun2018discovering}
A.~Y. Sun.
\newblock Discovering state-parameter mappings in subsurface models using
  generative adversarial networks.
\newblock {\em Geophysical Research Letters}, 45:11--137, 2018.

\bibitem{misra2019machine}
S.~Misra, H.~Li, and J.~He.
\newblock {\em Machine Learning for Subsurface Characterization}.
\newblock Gulf Professional Publishing, 2019.

\bibitem{camps2021deep}
G.~Camps-Valls, D.~Tuia, X.~X. Zhu, and M.~Reichstein.
\newblock {\em {Deep learning for the Earth sciences:~A comprehensive approach
  to remote sensing, climate science and geosciences}}.
\newblock John Wiley \& Sons, 2021.

\bibitem{ringner2008principal}
M.~Ringn{\'e}r.
\newblock What is principal component analysis?
\newblock {\em Nature Biotechnology}, 26(3):303--304, 2008.

\bibitem{bro2014principal}
R.~Bro and A.~K. Smilde.
\newblock Principal component analysis.
\newblock {\em Analytical Methods}, 6:2812--2831, 2014.

\bibitem{siler20213}
D.~L. Siler and J.~D. Pepin.
\newblock {3-D geologic controls of hydrothermal fluid flow at Brady geothermal
  field, Nevada, USA}.
\newblock {\em Geothermics}, 94:102112, 2021.

\bibitem{cichocki2009nonnegative}
A.~Cichocki, R.~Zdunek, A.~H. Phan, and S.-I Amari.
\newblock {\em Nonnegative Matrix and Tensor Factorizations:~Applications to
  Exploratory Multi-way Data Analysis and Blind Source Separation}.
\newblock John Wiley \& Sons, 2009.

\bibitem{vesselinov2019unsupervised}
V.~V. Vesselinov, M.~K. Mudunuru, S.~Karra, D.~O'Malley, and B.~S. Alexandrov.
\newblock Unsupervised machine learning based on non-negative tensor
  factorization for analyzing reactive-mixing.
\newblock {\em Journal of Computational Physics}, 395:85--104, 2019.

\bibitem{gonzalez2022monitoring}
K.~Gonzalez and S.~Misra.
\newblock Monitoring the $\mathrm{CO}_2$ plume migration during geological
  carbon storage using spatiotemporal clustering.
\newblock 2022.

\bibitem{shorten2019survey}
C.~Shorten and T.~M. Khoshgoftaar.
\newblock A survey on image data augmentation for deep learning.
\newblock {\em Journal of Big Data}, 6:1--48, 2019.

\bibitem{zhang2017mixup}
H.~Zhang, M.~Cisse, Yann~N Dauphin, and D.~L.-Paz.
\newblock {mixup:~Beyond empirical risk minimization}.
\newblock {\em arXiv preprint arXiv:1710.09412}, 2017.

\bibitem{jing2019neural}
Y.~Jing, Y.~Yang, Z.~Feng, J.~Ye, Y.~Yu, and M.~Song.
\newblock {Neural style transfer:~A review}.
\newblock {\em IEEE transactions on visualization and computer graphics},
  26:3365--3385, 2019.

\bibitem{tan2018survey}
C.~Tan, F.~Sun, T.~Kong, W.~Zhang, C.~Yang, and C.~Liu.
\newblock A survey on deep transfer learning.
\newblock In {\em International Conference on Artificial Neural Networks},
  pages 270--279, 2018.

\bibitem{finn2017model}
C.~Finn, P.~Abbeel, and S.~Levine.
\newblock Model-agnostic meta-learning for fast adaptation of deep networks.
\newblock In {\em International Conference on Machine Learning}, pages
  1126--1135, 2017.

\bibitem{vanschoren2018meta}
J.~Vanschoren.
\newblock {Meta-learning:~A survey}.
\newblock {\em arXiv preprint arXiv:1810.03548}, 2018.

\bibitem{liang2018understanding}
D.~Liang, F.~Yang, T.~Zhang, and P.~Yang.
\newblock Understanding mixup training methods.
\newblock {\em IEEE Access}, 6:58774--58783, 2018.

\bibitem{thulasidasan2019mixup}
S.~Thulasidasan, G.~Chennupati, J.~Bilmes, T.~Bhattacharya, and S.~Michalak.
\newblock {On mixup training:~Improved calibration and predictive uncertainty
  for deep neural networks}.
\newblock {\em arXiv preprint arXiv:1905.11001}, 2019.

\bibitem{zhang2020does}
L.~Zhang, Z.~Deng, K.~Kawaguchi, A.~Ghorbani, and J.~Zou.
\newblock How does mixup help with robustness and generalization?
\newblock {\em arXiv preprint arXiv:2010.04819}, 2020.

\bibitem{deng2021deep}
Y.~Deng, L.~Lu, L.~Aponte, A.~M. Angelidi, V.~Novak, G.~E. Karniadakis, and
  C.~S. Mantzoros.
\newblock Deep transfer learning and data augmentation improve glucose levels
  prediction in type 2 diabetes patients.
\newblock {\em NPJ Digital Medicine}, 4:1--13, 2021.

\bibitem{johnson2017pflotran}
T.~C. Johnson, G.~E. Hammond, and X.~Chen.
\newblock {PFLOTRAN-E4D:}~{A} parallel open source {PFLOTRAN} module for
  simulating time-lapse electrical resistivity data.
\newblock {\em Computers \& Geosciences}, 99:72--80, 2017.

\bibitem{ahmmed2020pflotran}
B.~Ahmmed, M.~K. Mudunuru, S.~Karra, S.~C. James, H.~Viswanathan, and J.~A.
  Dunbar.
\newblock {PFLOTRAN-SIP:~A PFLOTRAN Module for Simulating Spectral-Induced
  Polarization of Electrical Impedance Data}.
\newblock {\em Energies}, 13:6552, 2020.

\bibitem{hammond2012pflotran}
G.~E. Hammond, P.~C. Lichtner, C.~Lu, and R.~T. Mills.
\newblock {PFLOTRAN:~{R}eactive flow \& transport code for use on laptops to
  leadership-class supercomputers}.
\newblock {\em Groundwater Reactive Transport Models}, pages 141--159, 2012.

\bibitem{Hammondetal2014}
G.~E. Hammond, P.~C. Lichtner, and R.~T. Mills.
\newblock Evaluating the performance of parallel subsurface simulators:~{A}n
  illustrative example with {PFLOTRAN}.
\newblock {\em Water Resources Research}, 50:208--228, 2014.

\bibitem{lichtner2015pflotran}
P.~C. Lichtner, G.~E. Hammond, C.~Lu, S.~Karra, G.~Bisht, B.~Andre, R.~Mills,
  and J.~Kumar.
\newblock {PFLOTRAN User Manual:~A Massively Parallel Reactive Flow and
  Transport Model for Describing Surface and Subsurface Processes}.
\newblock Technical report, 2015.

\bibitem{johnsonetal2010}
T.~C. Johnson, R.~J. Versteeg, A.~Ward, F.~D. Day-Lewis, and A.~Revil.
\newblock Improved hydrogeophysical characterization and monitoring through
  parallel modeling and inversion of time-domain resistivity and
  induced-polarization data.
\newblock {\em Geophysics}, 75(4):WA27--WA41, 2010.

\bibitem{johnson2015four}
T.~Johnson, R.~Versteeg, J.~Thomle, G.~Hammond, X.~Chen, and J.~Zachara.
\newblock {Four-dimensional electrical conductivity monitoring of stage-driven
  river water intrusion:~Accounting for water table effects using a transient
  mesh boundary and conditional inversion constraints}.
\newblock {\em Water Resources Research}, 51:6177--6196, 2015.

\bibitem{rubin2003applied}
Y.~Rubin.
\newblock {\em Applied Stochastic Hydrogeology}.
\newblock Oxford University Press, 2003.

\bibitem{rubin2010bayesian}
Y.~Rubin, X.~Chen, H.~Murakami, and M.~Hahn.
\newblock A {B}ayesian approach for inverse modeling, data assimilation, and
  conditional simulation of spatial random fields.
\newblock {\em Water Resources Research}, 46, 2010.

\bibitem{scikit-learn}
F.~Pedregosa, G.~Varoquaux, A.~Gramfort, V.~Michel, B.~Thirion, O.~Grisel,
  M.~Blondel, P.~Prettenhofer, R.~Weiss, V.~Dubourg, J.~Vanderplas, A.~Passos,
  D.~Cournapeau, M.~Brucher, M.~Perrot, and E.~Duchesnay.
\newblock Scikit-learn: Machine learning in {P}ython.
\newblock {\em Journal of Machine Learning Research}, 12:2825--2830, 2011.

\bibitem{verma2018manifold}
V.~Verma, A.~Lamb, C.~Beckham, A.~Courville, I.~Mitliagkis, and Y.~Bengio.
\newblock {Manifold mixup:~Encouraging meaningful on-manifold interpolation as
  a regularizer}.
\newblock {\em arXiv preprint arXiv:1806.05236}, 7, 2018.

\bibitem{tf-keras}
{Keras API -- The high-level API of Tensorflow}.
\newblock \url{https://keras.io/},
  \url{https://www.tensorflow.org/api_docs/python/tf/keras}, 2021.
\newblock Accessed on:~2021-08-21.

\bibitem{omalley2019kerastuner}
T.~O'Malley, E.~Bursztein, J.~Long, F.~Chollet, H.~Jin, and L.~Invernizzi.
\newblock {Keras Tuner}.
\newblock \url{https://github.com/keras-team/keras-tuner}, 2019.

\bibitem{dalcin2021mpi4py}
L.~Dalcin and Y.-.~L. Fang.
\newblock {Mpi4py:~Status update after 12 years of development}.
\newblock {\em Computing in Science \& Engineering}, 23:47--54, 2021.

\bibitem{palach2014parallel}
J.~Palach.
\newblock {\em Parallel Programming with Python}.
\newblock Packt Publishing Ltd, 2014.

\bibitem{NERSC2021}
{NERSC -- National Energy Research Scientific Computing Center}.
\newblock \url{https://www.nersc.gov/}, 2021.
\newblock Accessed: 2021-08-01.

\bibitem{rand_seed_popular_1}
S.~Seemayer.
\newblock {The most popular random seeds}.
\newblock
  \url{{https://blog.semicolonsoftware.de/the-most-popular-random-seeds/}},
  2015.
\newblock Accessed on:~2021-08-21.

\bibitem{rand_seed_popular_2}
A.~Bilogur.
\newblock {Most Common Random Seeds -- Kaggle}.
\newblock \url{{https://www.kaggle.com/residentmario/kernel16e284dcb7}}, 2017.
\newblock Accessed on:~2021-08-21.

\bibitem{rand_seed_popular_3}
A.~Bilogur.
\newblock {What are the most popular random seeds?}
\newblock \url{{https://www.residentmar.io/2016/07/08/randomly-popular.html}},
  2016.
\newblock Accessed on:~2021-08-21.

\bibitem{NERSC-Cori}
{NERSC Cori compute nodes}.
\newblock \url{https://docs.nersc.gov/systems/cori/#haswell-compute-nodes},
  2021.
\newblock Accessed: 2021-08-01.

\bibitem{seabold2010statsmodels}
S.~Seabold and J.~Perktold.
\newblock {Statsmodels:~Econometric and statistical modeling with python}.
\newblock In {\em Proceedings of the 9th Python in Science Conference},
  volume~57, page~61, 2010.

\bibitem{garcia2015data}
S.~Garc{\'\i}a, J.~Luengo, and F.~Herrera.
\newblock {\em Data preprocessing in data mining}, volume~72.
\newblock Springer, 2015.

\bibitem{smith2018less}
J.~S. Smith, B.n Nebgen, N.~Lubbers, O.~Isayev, and A.~E. Roitberg.
\newblock Less is more:~sampling chemical space with active learning.
\newblock {\em The Journal of Chemical Physics}, 148:241733, 2018.

\bibitem{kim2020puzzle}
J.-H. Kim, W.~Choo, and H.~O. Song.
\newblock {Puzzle mix:~Exploiting saliency and local statistics for optimal
  mixup}.
\newblock In {\em International Conference on Machine Learning}, pages
  5275--5285, 2020.

\bibitem{hwang2021mixrl}
S.-H. Hwang and S.~E. Whang.
\newblock {MixRL:~Data mixing augmentation for regression using reinforcement
  learning}.
\newblock {\em arXiv preprint arXiv:2106.03374}, 2021.

\bibitem{wang2019few}
Y.~Wang and Q.~Yao.
\newblock {Few-shot learning:~A survey}.
\newblock 2019.

\bibitem{ganaie2021ensemble}
M.~A. Ganaie and M.~Hu.
\newblock {Ensemble deep learning:~A review}.
\newblock {\em arXiv preprint arXiv:2104.02395}, 2021.

\bibitem{raissi2019physics}
M.~Raissi, P.~Perdikaris, and G.~E. Karniadakis.
\newblock {Physics-informed neural networks:~A deep learning framework for
  solving forward and inverse problems involving nonlinear partial differential
  equations}.
\newblock {\em Journal of Computational Physics}, 378:686--707, 2019.

\bibitem{li2020fourier}
Z.~Li, N.~Kovachki, K.~Azizzadenesheli, B.~Liu, K.~Bhattacharya, A.~Stuart, and
  A.~Anandkumar.
\newblock Fourier neural operator for parametric partial differential
  equations.
\newblock {\em arXiv preprint arXiv:2010.08895}, 2020.

\bibitem{karra2021adjointnet}
S.~Karra, B.~Ahmmed, and M.~K. Mudunuru.
\newblock {AdjointNet:~Constraining machine learning models with physics-based
  codes}.
\newblock {\em arXiv preprint arXiv:2109.03956}, 2021.

\end{thebibliography}

%
\end{document}